\documentclass[prd,twocolumn,superscriptaddress,floatfix,nopacs,preprintnumbers,nofootinbib]{revtex4-2}
\usepackage[utf8]{inputenc}

\usepackage[caption = false]{subfig}
\usepackage[normalem]{ulem}
\usepackage{xcolor}

\usepackage{mathrsfs}
\usepackage{amssymb,bm}
\usepackage{amsmath}
\usepackage{mathtools}
\usepackage{slashed}
\usepackage{physics}

\usepackage{tikz}

\newcommand{\nc}{N_\mathrm{c}}
\newcommand{\aem}{\alpha_\mathrm{em}}
\newcommand{\gev}{\mathrm{GeV}}
\newcommand{\jpsi}{$\mathrm{J}/\psi$ }
\newcommand{\jpsim}{\mathrm{J}/\psi}
\newcommand{\bt}{\mathbf{b}}
\newcommand{\rt}{\mathbf{r}}
\newcommand{\Bt}{\mathbf{B}}
\newcommand{\xpom}{x_\mathbb{P}}
\newcommand{\pt}{\mathbf{p}}
\newcommand{\xt}{\mathbf{x}}

\newcommand{\kt}{\mathbf{k}}
\newcommand{\yt}{\mathbf{y}}
\newcommand{\Deltat}{\mathbf{\Delta}}
\newcommand{\Mcal}{\mathcal{M}}
\newcommand{\Fcal}{\mathcal{F}}
\newcommand{\Acal}{\mathcal{A}}

\newcommand{\lqcd}{\Lambda_\text{QCD}}
\newcommand{\nf}{N_\mathrm{f}}

\usepackage[breaklinks,colorlinks,citecolor=citcolor,urlcolor=blue,linkcolor=lcolor]{hyperref}

\definecolor{lcolor}{rgb}{0.5,0,0}
\definecolor{citcolor}{rgb}{0,0.3,0.0}

\newcommand{\as}{\alpha_\mathrm{s}}
\newcommand{\der}{\mathrm{d}}

\begin{document}

\author{Heikki M\"{a}ntysaari}
\email{heikki.mantysaari@jyu.fi}
\affiliation{
Department of Physics, University of Jyväskylä,  P.O. Box 35, 40014 University of Jyväskylä, Finland
}
\affiliation{
Helsinki Institute of Physics, P.O. Box 64, 00014 University of Helsinki, Finland
}
\author{Farid Salazar}
\email{salazar@physics.ucla.edu}
\affiliation{Department of Physics and Astronomy, University of California, Los Angeles, CA 90095, USA}
\affiliation{Mani L. Bhaumik Institute for Theoretical Physics, University of California, Los Angeles, CA 90095, USA}
\affiliation{Nuclear Science Division, Lawrence Berkeley National Laboratory, Berkeley, CA 94720, USA}
\affiliation{Physics Department, University of California, Berkeley, CA 94720, USA}
\author{Bj\"{o}rn Schenke}
\email{bschenke@bnl.gov}
\affiliation{Physics Department, Brookhaven National Laboratory, Bldg. 510A, Upton, NY 11973, USA}

\title{Nuclear geometry at high energy from exclusive vector meson production  }

\begin{abstract}
We show that when saturation effects are included one obtains a good description of the exclusive \jpsi production  spectra in ultra peripheral lead-lead collisions as recently measured by the ALICE Collaboration at the LHC. As exclusive spectra are sensitive to the spatial distribution of nuclear matter at small Bjorken-$x$, 
this implies that gluon saturation effects modify the impact parameter profile of the target as we move towards small $x$. In addition to saturation effects, we find a preference for larger nuclear strong-interaction radii compared to the typical charge radius. 
We demonstrate the role of finite photon transverse momentum and the interference between the cases for which the role of photon emitter and target are switched between the nuclei.
We show that these effects are comparable to the experimental precision for $p_T$-differential cross sections and as such need to be included when comparing to LHC data.
Finally, the integrated \jpsi production cross sections from the LHC and preliminary transverse momentum spectra from RHIC are shown to prefer calculations with fluctuating nucleon substructure, although these datasets would require even stronger saturation effects than predicted from our framework.

\end{abstract}

\maketitle

\section{Introduction}

Exclusive particle production processes in Deep Inelastic Scattering (DIS) are powerful probes of the structure of protons and nuclei at high energy. The exclusive nature of the process ensures that there is no net color charge transferred from the target, which means that at least two gluons need to be exchanged. This renders the cross section approximately proportional to the \emph{square} of the gluon distribution at leading order~\cite{Ryskin:1992ui} (at next-to-leading order the relation is less direct~\cite{Eskola:2022vpi}). Additionally, measuring the total momentum transfer to the target is possible by measuring the produced particle, e.g. a vector meson. As the momentum transfer is the Fourier conjugate to the impact parameter, exclusive processes provide access to the spatial distribution of nuclear matter in  protons and nuclei. Indeed, multi dimensional imaging using exclusive photon or vector meson production processes is a central part of the physics programs of future nuclear-DIS facilities, including the EIC~\cite{AbdulKhalek:2021gbh,Aschenauer:2017jsk}, LHeC/FCC-he~\cite{Agostini:2020fmq} and EicC~\cite{Anderle:2021wcy}.

Before these future facilities are realized, it is also possible to study exclusive vector meson production at high energy in the photoproduction region in Ultra Peripheral Collisions (UPCs) at RHIC and at the LHC~\cite{Bertulani:2005ru,Klein:2019qfb}. In UPCs the impact parameter is so large that strong interactions are suppressed, and one of the colliding nuclei acts as a source of quasi real photons, which probe the other nucleus. In particular, ultra peripheral heavy ion collisions provide access to photon-nucleus scattering at collider energies for the first time. 

Experiments at both RHIC and at the LHC have performed first measurements of the exclusive \jpsi photoproduction cross section in heavy ion  UPCs~\cite{ALICE:2012yye,ALICE:2013wjo,ALICE:2014eof,ALICE:2018oyo,ALICE:2021gpt,ALICE:2019tqa,ALICE:2021tyx,LHCb:2014acg, LHCb:2018rcm,LHCb:2022ahs,LHCb:2021bfl,LHCb:2022ahs,CMS:2016itn,PHENIX:2009xtn}. These measurements have been extensively studied in the context of saturation physics, e.g.~in Refs.~\cite{Sambasivam:2019gdd,Lappi:2013am,Cepila:2017nef,Mantysaari:2017dwh,Goncalves:2017wgg,Bendova:2020hbb} (see also Refs.~\cite{Toll:2012mb,Mantysaari:2019jhh,Lappi:2010dd,Caldwell:2010zza} where vector meson production in photon-nucleus collisions is studied).
Very recently first measurements differential in the meson transverse momentum $\pt$ or squared momentum transfer $|t|$ have also become available~\cite{LHCb:2022ahs,ALICE:2021tyx, star}. These new developments make it possible to study the geometric structure of nuclei, including event-by-event fluctuations~\cite{Mantysaari:2020axf}, in a so far unexplored kinematical domain down to $x \sim 10^{-5}$. This possibility is the main motivation behind this work.

We calculate within the Color Glass Condensate framework~\cite{Iancu:2003xm,Albacete:2014fwa,Blaizot:2016qgz,Gelis:2010nm,Morreale:2021pnn} exclusive \jpsi production in ultra peripheral lead-lead and gold-gold collisions. In particular we show how the non-linear saturation effects change the nuclear geometry (as measured by the \jpsi spectra) when one moves from the low-energy region described in terms of nucleon positions following the nuclear density distribution, such as the Woods-Saxon distribution \,\cite{Woods:1954zz}, to the region of strong color fields in the small momentum fraction $x$ region probed in collider experiments. Compared to our previous study~\cite{Mantysaari:2017dwh} we use a full CGC based setup including perturbative small-$x$ evolution calculated by solving the JIMWLK equation (see e.g.~\cite{Mueller:2001uk}), which also describes the geometry evolution \cite{Schlichting:2014ipa}. Additionally we take into account the interference effect due to the fact that it is not possible to know which nucleus emitted the photon \cite{Klein:1999gv,Bertulani:2005ru} and the non-zero photon transverse momentum~\cite{Xing:2020hwh}.

This manuscript is organized as follows. In Sec.~\ref{sec:upc} we discuss how ultra peripheral collisions can be considered as photon-nucleus events, and show how the interference effect and photon transverse momentum are taken into account in our calculations. The calculation of exclusive vector meson production from a CGC setup including the small-$x$ evolution is presented in Sec.~\ref{sec:vm_production}. Numerical results compared to  LHC and RHIC data are presented in Sec.~\ref{sec:results} before we present our conclusions in Sec.~\ref{sec:conclusions}.

\section{Ultra peripheral collisions}
\label{sec:upc}
There are two indistinguishable contributions to the exclusive vector meson production in ultra peripheral collisions, as both of the colliding nuclei can act as a photon source. Consequently there is also a quantum mechanical interference contribution which becomes important at small vector meson transverse momentum $|\pt| $~\cite{Bertulani:2005ru}. Additionally, although the photons are quasi real with their virtuality limited by the nuclear size $Q^2 \lesssim 1/R_A^2$, they carry a non-zero transverse momentum that can have an effect on the vector meson transverse momentum spectra, especially near diffractive minima. 

In order to include both the photon transverse momentum $\kt$ (which is related to transverse distance between the nuclei $\Bt$ via Fourier transform) and the interference contribution, we follow Ref.~\cite{Xing:2020hwh}. Let us first consider coherent vector meson production, where the target remains intact and one averages the scattering amplitude over the target configurations $\Omega$~\cite{Good:1960ba,Caldwell:2009ke}. The result derived in Ref.~\cite{Xing:2020hwh} can be written as (see Appendix~\ref{appendix:amplitude} for details)
\begin{align}
\label{eq:full_xs}
    \frac{\der \sigma^{A_1 + A_2 \to V + A_1 + A_2}}{\der \pt^2 \der y }= \frac{1}{4\pi}  \int_{|\Bt|>B_{\mathrm{ min}}} \!\!\!\!\!\!\!\!\!\!\!\!\!\! \der^2 \Bt |\langle \Mcal^j(y,\pt,\Bt) \rangle_{\Omega}|^2 \,,
\end{align}
where
\begin{align}
    &\Mcal^j(y,\pt,\Bt) =  \Bt^j \Mcal_0(y,\pt,\Bt) - \Mcal^j_1(y,\pt,\Bt)  \nonumber \\
    &- \left[\Bt^j \Mcal_0(-y,\pt,-\Bt)  +  \Mcal^j_1(-y,\pt,-\Bt) \right] e^{-i\pt\cdot\Bt} \,, \label{eq:Mi_simple}
\end{align}
and
\begin{align}
    \Mcal_{0}(y,\pt,\Bt) &= \int \der^2 \bt e^{-i\pt\cdot\bt} (-i\widetilde{\Acal}(y,\bt)) \widetilde{\Fcal}_{S}(y,\bt-\Bt) \,, \nonumber \\
    \Mcal_{1}^j(y,\pt,\Bt) &= \int \der^2 \bt e^{-i\pt\cdot\bt} (-i\widetilde{\Acal}(y,\bt))  \bt^j \widetilde{\Fcal}_{S}(y,\bt-\Bt) \, .
\end{align}
An equivalent expression to Eq.\,\eqref{eq:Mi_simple} is given in Eq.\,\eqref{eq:Amplitude_coordinate-space2} where the symmetry in the exchange between photon emitter and target is manifest.

The vector meson production amplitude in photon-target interaction, $\widetilde{A}(y,\bt)$,  is discussed in more detail in Sec.\,\ref{sec:vm_production}.

The transverse coordinate index is $j=1,2$.
Here the vector meson $V$ rapidity is denoted by $y$, and its transverse momentum $\pt$ is obtained as a vector sum of the photon transverse momentum $\kt$ and the nuclear momentum transfer $\Deltat$. The photon transverse momentum and the momentum transfer are not explicitly visible above as we work in coordinate space, see discussion in Appendix~\ref{appendix:amplitude}. 

The impact parameter of the photon-nucleus collision is denoted by $\bt$. The integral over the transverse separation between the two nuclei, $\Bt$, is limited from below in ultra peripheral collisions, and we use $B_\text{min}=2R_A$ where $R_A=6.62$ fm for Pb and $R_A=6.37$ fm for Au unless stated otherwise. 

The function $\Fcal_S$ describes the electromagnetic field of the nucleus calculated using an equivalent photon approximation Fourier transformed into coordinate space. As we only need the electromagnetic field at distances $|\Bt| \ge 2R_A$, the nuclear form factor can be replaced by that of a point particle following Gauss' law (we have confirmed that using a Woods-Saxon form factor has negligible effect on our results). In this case the function $\Fcal_S$ reads
\begin{equation}
\label{eq:point_charge_field}
    \widetilde{\Fcal}_S(y, \Bt) 
     = \frac{Z {\aem}^{1/2} \omega}{\pi \gamma}  \frac{1}{|\Bt|} K_1\left( \frac{\omega |\Bt|}{\gamma} \right) \,. 
\end{equation}
The photon energy is $\omega = (M_V/2) e^{y}$, $Z$ is the ion charge and $\gamma=A\sqrt{s}/(2M_A)$ where $M_A$ is the mass of the nucleus. The vector meson mass is denoted by $M_V$. As discussed above and in Appendix~\ref{appendix:amplitude}, the impact parameter $\Bt$ is related to the photon transverse momentum and as such the size of the nucleus sets the scale for the photon transverse momenta. In this work we use a sharp cutoff $|\Bt|>B_\text{min}$ which potentially has an effect on the photon $k_T$ distribution as discussed in Ref.~\cite{Klein:2020jom}.

The results shown in this work are not highly sensitive to the $B_\text{min}$ cut: for example the total coherent \jpsi production cross section at the LHC discussed in Sec.~\ref{sec:tint_xs} changes by $\sim 3\%$ when the minimum distance is changed by $10\%$.

We further note that at midrapidity and for coherent production (using the fact that $\langle -i\widetilde{\Acal}(y,\bt) \rangle_{\Omega}$ is real) the amplitude in Eq.\,\eqref{eq:Mi_simple} averaged over configurations can be cast into a simple form
\begin{align}
    &\langle \Mcal^j(0,\pt,\Bt) \rangle_{\Omega}=  2 i e^{i\pt\cdot\Bt/2} \nonumber \\
    &\times \mathrm{Im} \Big \{ e^{-i\pt\cdot\Bt/2} \left[\Bt^j \langle \Mcal_0(0,\pt,\Bt) \rangle_{\Omega} - \langle \Mcal^j_1(0,\pt,\Bt) \rangle_{\Omega} \right]  \Big \}\,.
\end{align}

Let us next discuss some commonly used approximations. First, as the photon transverse momentum is small, $\kt^2 \lesssim Q^2 \sim 1/R_A^2$, it can usually (but not around diffractive minima) be neglected. In coordinate space this corresponds to assuming $|\Bt| \gg |\bt|$. In this case the cross section can be written as

\begin{widetext}
    \begin{align}
\label{eq:interference_only}
    &\frac{\dd[] \sigma^{A_1+A_2 \to V+A_1+A_2}}{\dd[]{\pt^2} \dd{y}} = \frac{1}{4\pi}  \Bigg\{ 
    N(\omega_+) \left| \int \dd[2]{\bt} e^{-i\pt\cdot\bt} \langle -i \widetilde{\Acal}(y,\bt) \rangle_\Omega \right|^2 
    + N(\omega_-) \left| \int \dd[2]{\bt} e^{-i\pt\cdot\bt} \langle -i \widetilde{\Acal}(-y,\bt) \rangle_\Omega \right|^2 \nonumber \\
    &-  \left[\int_{|\Bt|>B_\text{min}} \!\!\!\!\!\!\!\!\!\!\! \dd[2]{\Bt} \sqrt{n(\omega_-,\Bt)n(\omega_+,\Bt)} \cos (\pt \cdot \Bt )\right] 
    2 \mathrm{Re}\left[\left(\int \dd[2]{\bt} e^{-i\pt\cdot\bt} \langle -i\widetilde{\Acal}(y,\bt)\rangle_\Omega \right)  \left(\int \dd[2]{\bt} e^{i\pt\cdot\bt} \langle -i\widetilde{\Acal}(-y,\bt)\rangle_\Omega \right)\right] \Bigg\},
\end{align}
\end{widetext}
where
$\omega_\pm = (M_V/2) e^{\pm y}$ and we write the photon flux as
\begin{equation}
\label{eq:bint_flux}
N(\omega_\pm) = \int_{|\Bt|>B_\text{min}} \!\!\!\!\!\!\! \dd[2]{\Bt} n(\omega_\pm, \Bt) \,,
\end{equation}
and
\begin{equation}
\label{eq:flux}
    n(\omega,\Bt) = |\Bt|^2 \widetilde{\Fcal}_S(y,|\Bt|)^2 = \frac{Z^2 \aem \omega^2}{\pi^2 \gamma^2}  K_1^2\left( \frac{\omega |\Bt|}{\gamma}\right).
\end{equation}
At midrapidity this further simplifies and one obtains
\begin{multline}
\label{eq:interference_only_midrapidity}
    \left.\frac{\dd[] \sigma^{A_1+A_2 \to V+A_1+A_2}}{\dd[]{\pt^2} \dd{y}}\right|_{y=0} = \frac{1}{4\pi} \int_{|\Bt|>B_\text{min}} \!\!\!\!\!\!\!\!\!\!\! \dd[2]{\Bt} \\
    \times \left| \int \dd[2]{\bt}  e^{-i\pt\cdot\bt} \langle -i \widetilde{\Acal}(0,\bt)\rangle_\Omega \right|^2 \\
    \times 2 n(\omega, \Bt) \left[1 - \cos (\pt \cdot \Bt )  \right],
\end{multline}
This is the result derived also in Ref.~\cite{Bertulani:2005ru}, and it is now clear that the interference effect results in a cross section that vanishes at $\pt=0$ at midrapidity.

Furthermore, if both the interference effect and the photon transverse momentum are neglected, one recovers the standard result
\begin{multline}
\label{eq:dsigma_dy_nointerf_nokt}
   \frac{\der \sigma^{A_1+A_2 \to V+A_1+A_2}}{\dd \pt^2 \dd{y}} =  N(\omega_{+}) \frac{\der \sigma_{+}^{\gamma^* + A \to V + A}}{\der \pt^2}  \\
   +  N(\omega_{-}) \frac{\der \sigma_{-}^{\gamma^* + A \to V + A}}{\der \pt^2} ,
\end{multline}
where the diffractive cross section for the $\gamma^* + A \to V +A$ subprocess reads
\begin{equation}
\label{eq:gamma_A_xs}
    \frac{\der \sigma_{\pm}^{\gamma^* + A \to V + A}}{\dd \pt^2} = \frac{1}{4\pi} \left| \int \dd[2]{\bt} e^{-i\pt\cdot\bt} \left\langle-i\widetilde{\Acal}(\pm y,\bt)\right\rangle_\Omega \right|^2.
\end{equation}
Here the squared center-of-mass energy for the photon-nucleon system $W^2$ determines $y$, and $\der \sigma_{+}^{\gamma^* + A \to V + A}$ and $\der \sigma_{-}^{\gamma^* + A \to V + A}$ refer to the photon-nucleus cross sections where the target structure is probed at different longitudinal momentum fractions $\xpom = (M_V/\sqrt{s})e^{\mp y}$.

Let us finally discuss incoherent vector meson production, referring to the events where the target nucleus is excited (denoted by $A^*$) and dissociates. In that case the average over target configurations $\Omega$ is taken at the cross section level to calculate the total diffractive vector meson production cross section, from which the coherent cross section is subtracted~\cite{Caldwell:2009ke}. Thus the cross section becomes sensitive to the event-by-event fluctuations in the scattering amplitude, and probes the target spatial density fluctuations~\cite{Mantysaari:2016ykx,Mantysaari:2016jaz,Mantysaari:2020axf}.

The incoherent cross section dominates at large $\pt^2 \gtrsim 1/R_A^2$, where $R_A$ is the size of the nucleus. 
Because the photon transverse momentum is limited by the inverse size of the nucleus, we neglect the photon $\kt$ when calculating the incoherent cross section. The interference effect also has a non-negligible contribution only in the very small $\pt^2$ region as demonstrated in Appendix~\ref{appendix:interference}. As this is exactly the region where the coherent process dominates, the interference effect is  negligible for the incoherent cross section. 

The incoherent cross section in $\gamma^* + A$ scattering can be written as
\begin{multline}
\label{eq:gamma_A_xs_incoh}
    \frac{\der \sigma_{\pm}^{\gamma^* + A \to V + A^*}}{\dd \pt^2 } = \frac{1}{4\pi} \left\langle \left| \int \dd[2]{\bt} e^{-i\pt\cdot\bt} (-i \widetilde{\Acal}(\pm y,\bt)) \right|^2 \right\rangle_\Omega \\
    - \frac{1}{4\pi} \left| \int \dd[2]{\bt} e^{-i\pt\cdot\bt} \left\langle-i\widetilde{\Acal}(\pm y,\bt)\right\rangle_\Omega \right|^2.
\end{multline}
The incoherent cross section in ultra peripheral collisions is then obtained using Eq.~\eqref{eq:interference_only} (or Eq.~\eqref{eq:dsigma_dy_nointerf_nokt} which is a very good approximation in the kinematical domain where the incoherent process is relevant).
As a variance the incoherent cross section is directly proportional to the amount of event-by-event fluctuations in the scattering amplitude. Additionally, as the impact parameter is the Fourier conjugate to the momentum transfer, the incoherent cross section in different transverse momentum regions probes these fluctuations at different length scales as we will demonstrate in Sec.~\ref{sec:results}, see also Refs.~\cite{Demirci:2022wuy,Mantysaari:2019jhh,Mantysaari:2017dwh,Lappi:2010dd}.

\section{Vector meson production at high energy}
\label{sec:vm_production}
At high energies it is convenient to describe vector meson production in photon-nucleus scattering in the dipole picture. In the frame where the photon has a large longitudinal momentum the photon splits into a quark-antiquark pair long before it interacts with the color field of the target. The $\gamma\to q\bar q$ splitting is a QED process and described in terms of the photon light front wave function $\Psi_\gamma$~\cite{Kovchegov:2012mbw}. The elastic dipole-target interaction is given in terms of the dipole-target scattering amplitude $N(\rt,\bt,z,\xpom)$, where $z$ is the fraction of the photon light-cone momentum carried by the quark, $\rt$ is the quark-antiquark separation and $\bt$ the impact parameter (distance from the center of the nucleus to the center-of-mass of the dipole\footnote{More precisely, we have $\bt = z \xt + (1-z)\yt$, where $\xt$ and $\yt$ are the transverse positions of the quark and anti-quark respectively.}), which depends implicitly on the target color charge configuration $\Omega$. Finally, a non-perturbative vector meson wave function $\Psi_V$ is used to describe the formation of a vector meson after the interaction with the target. The scattering amplitude for this process at leading order reads~\cite{Kowalski:2006hc,Hatta:2017cte}
\begin{align}
     -i\widetilde{\mathcal{A}}(y,\bt) = \int \dd[2]{\rt}  \int_0^1 \frac{\dd{z}}{4\pi}  &[\Psi_V^* \Psi_\gamma](Q^2,\rt,z) \nonumber \\
     &\times N(\rt,\bt,z,\xpom) \,,
\end{align}
at fixed impact parameter $\bt$.
Here we take $Q^2=0$ for the photon virtuality. 
We only include the contribution where the photon and vector meson are transversely polarized, as the polarization changing contribution is negligible in the kinematical domain studied in this work~\cite{Mantysaari:2020lhf}.

We use the Boosted Gaussian parametrization for the \jpsi wave function $\Psi_V$ with the parameters given in Ref.~\cite{Mantysaari:2018nng}. There are also other vector meson light front wave functions proposed in the literature, e.g.~in Refs.~\cite{Lappi:2020ufv,Li:2017mlw,Li:2021cwv}. Using other wave function models mostly affects the overall normalization of the cross section and has only a minor effect on the shape of the $t$ spectra, which is the main focus of this work. Additionally, the nucleon density, which controls this normalization, is fixed to reproduce the \jpsi photoproduction cross section at HERA as will be discussed shortly. Consequently, our results are expected to depend only weakly on the actual model chosen for the \jpsi wave function. We also note that the field is rapidly moving towards next-to-leading order accuracy, in particular the exclusive vector meson production cross section has recently become available at NLO~\cite{Mantysaari:2022kdm,Mantysaari:2022bsp,Mantysaari:2021ryb,Escobedo:2019bxn,Boussarie:2016bkq}.
However, our main focus is on the role of saturation effects on nuclear geometry, using the very generic relation between the impact parameter and the momentum transfer, and we expect the NLO corrections to have only a moderate effect on our results. For example the higher order corrections have been shown to have only a few percent effect on the nuclear suppression factor in exclusive vector meson production~\cite{Lappi:2021oag}.

To describe the dipole-target scattering and calculate the dipole amplitude $N(\rt,\bt,z,\xpom)$ 
we use a Color Glass Condensate based framework as in Ref.~\cite{Mantysaari:2020lhf} (see also Refs.~\cite{Mantysaari:2019jhh,Mantysaari:2018zdd,Mantysaari:2016ykx,Mantysaari:2016jaz}). The target structure is described in terms of the Wilson lines $V(\xt)$ (that depend on the target configuration $\Omega$ and the target momentum fraction $\xpom$), and the dipole-target scattering amplitude reads\footnote{Note that by including the $z$ dependence in the argument of the dipole, we are effectively including the non-forward phase \cite{Hatta:2017cte}.}
\begin{align}
    &N(\rt,\bt,z,\xpom)  \nonumber \\
    & = 1 - \frac{1}{\nc} \tr \left[ V\left(\bt + (1-z)\rt\right) V^\dagger\left(\bt - z\rt\right) \right]. 
\end{align}
To obtain the Wilson lines that describe the target structure at $\xpom=0.01$, we use the McLerran-Venugopalan model~\cite{McLerran:1993ka,McLerran:1993ni} where the color charge density $\rho$ is assumed to be  a local Gaussian variable (see also Refs.~\cite{Dumitru:2020gla,Dumitru:2021tvw} where a complementary approach is taken to determine the proton color charge correlator):
\begin{multline}
    g^2 \langle \rho^a(x^-,\xt) \rho^b(y^-,\yt)\rangle = \delta^{ab} \delta^{(2)}(\xt-\yt) \delta(x^- - y^-) \\
    \times g^4 \lambda_A(x^-).
\end{multline}

The local color charge density $\mu^2 = \int \dd{x^-} \lambda_A(x^-)$ is determined from the local saturation scale $Q_s^2$ of the target extracted from the IPsat parametrization~\cite{Kowalski:2003hm} fitted to the HERA data~\cite{Rezaeian:2012ji}. The proportionality constant $c$ in the relation 
\begin{equation} 
\label{eq:qsmu}
Q_s = c g^2\mu\,,
\end{equation}
is determined by requiring a correct normalization to the HERA \jpsi production data as discussed below (see also Ref.~\cite{Lappi:2007ku}). In the IPsat model the local saturation scale $Q_s^2(\xt)$ is proportional to the local transverse density $T_p(\xt)$.  For nuclei, we first sample nucleon positions from a Woods-Saxon distribution, and then calculate the total density by summing the nucleon density profiles~\cite{Schenke:2012wb}.

The proton geometry is constrained by HERA data. When nucleon substructure is not included, the nucleon density profile is approximated by a Gaussian
\begin{equation}
    T_p(\bt) = \frac{1}{2\pi B_p} e^{-\bt^2/(2B_p)}.
\end{equation}
We also calculate results including the nucleon shape fluctuations following Refs.~\cite{Mantysaari:2016ykx,Mantysaari:2016jaz}, in which case we have
\begin{equation}
    T_p(\bt) = \frac{1}{2\pi B_q N_q} \sum_{i=1}^{N_q} p_i e^{-(\bt - \bt_i)^2/(2B_q)},
\end{equation}
where the hot spot positions $\bt_i$ are sampled from a Gaussian distribution that has a width $B_{qc}$, and the center-of-mass is moved to the origin after the sampling. In this work we use $N_q=3$ hot spots. The factors $p_i$ are used to implement additional density ($Q_s^2$) fluctuations, and are sampled from a log-normal distribution
\begin{equation}
\label{eq:qsfluct}
P\left( \ln p_i \right) = \frac{1}{\sqrt{2\pi}\sigma} \exp \left[- \frac{\ln^2 p_i}{2\sigma^2}\right].
\end{equation}
The sampled $p_i$ are normalized by the expectation value of the distribution  $E[p_i]=e^{\sigma^2/2}$ in order to keep the average density unmodified. 

The Wilson lines are obtained in terms of the sampled color charge configuration as a path ordered exponential:
\begin{equation}
    V(\xt) = P_- \exp \left\{ -ig \int \dd{x^-} \frac{\rho^a(x^-,\xt) t^a}{\nabla^2_{\xt} - \tilde m^2} \right\},
\end{equation}
where $\tilde m$ is an infrared regulator whose value is also fixed by the HERA data.

To obtain the Wilson lines, and the dipole-target amplitude, at smaller $\xpom<0.01$ we evolve the sampled configurations event-by-event by solving the JIMWLK equation with running coupling corrections following again Refs.~\cite{Lappi:2013zma,Mantysaari:2018zdd,Mantysaari:2020lhf} (see also Ref.~\cite{Cali:2021tsh}). The long distance Coulomb tails are regulated by introducing an exponential suppression for the gluon emission at long distances to the JIMWLK kernel by replacing
\begin{equation}
    K^i_{\xt} = \frac{x^i}{\xt^2} \to m |\xt| K_1(m|\xt|) \frac{x^i}{\xt^2}.
\end{equation}
The strong coupling constant as a function of transverse distance scale $r$ reads
\begin{equation}
    \as(r) = \frac{12\pi}{(11\nc - 2\nf) \ln \left[ \left(\frac{\mu_0^2}{\lqcd^2}\right)^{1/\zeta} + \left(\frac{4}{r^2\lqcd^2}\right)^{1/\zeta} \right]^\zeta}\,.
\end{equation}
The value of the coordinate space $\lqcd$ is fixed by the energy dependence of the HERA \jpsi\ production data as we will discuss next, and we use $\mu_0=0.28\,\gev$, $\zeta=0.2$ and $\nf=3$ as e.g. in Refs.~\cite{Mantysaari:2018zdd,Lappi:2012vw}.

As alluded to several times already, the model parameters are constrained by comparing with the coherent and incoherent \jpsi photoproduction data measured at HERA at photon-proton center-of-mass energy $W=75\,\gev$~\cite{H1:2013okq} (when nucleon shape fluctuations are not included, we only require a good description of the coherent spectra). Additionally, we require that the total coherent \jpsi photoproduction cross section in $\gamma + p$ scattering as a function of $W$ is compatible with the H1~\cite{H1:2005dtp,H1:2013okq}, ZEUS~\cite{ZEUS:2002wfj}, ALICE~\cite{ALICE:2014eof,ALICE:2018oyo} and LHCb~\cite{LHCb:2014acg, LHCb:2018rcm} data.

For comparison, we also show some results obtained by using a dipole amplitude from the IPsat parametrization~\cite{Mantysaari:2018nng}. When the IPsat parametrization is used we also include the so called skewedness and real part corrections  calculated as in Ref.~\cite{Mantysaari:2017dwh}.

The model parameters and the numerical values determined are summarized below (see also Ref.~\cite{Mantysaari:2022ffw} for a recent Bayesian analysis of the proton shape fluctuations without the JIMWLK evolution).
\begin{itemize}
    \item Proportionality constant $c$ between the color charge density and saturation scale  in Eq.~\eqref{eq:qsmu}: $c=0.638$ with no proton shape fluctuations, $c=0.643$ with fluctuations. 
    \item Proton size at the initial condition $\xpom=0.01$: $B_p=3\,\gev^{-2}$ with no proton shape fluctuations, and $B_{qc}=3.3\,\gev^{-2}, B_q=0.3\,\gev^{-2}$ when the proton shape fluctuations are included.
    \item Magnitude of $Q_s$ fluctuations: $\sigma=0.7$ in Eq.~\eqref{eq:qsfluct} (used with fluctuating nucleon substructure).
    \item Infrared regulators $m = \tilde m = 0.4\,\gev $.
    \item Scale of the strong coupling constant in coordinate space: $\lqcd = 0.025\,\gev$ (without nucleon shape fluctuations), $\lqcd=0.040\,\gev$ (with shape fluctuations).
\end{itemize}
The value for the coordinate space $\lqcd$ (which is different from momentum space $\lqcd$~\cite{Kovchegov:2006vj,Lappi:2011ju,Lappi:2012vw}) may appear to be small. We note that generically the leading order (with $\as \ln 1/x$ contributions resummed by small-$x$ evolution equations) fits in the CGC setup result in too large evolution speed in Bjorken-$x$ when compared to HERA data~\cite{Albacete:2010sy,Lappi:2013zma,Mantysaari:2018zdd}. In order to obtain an $x$ dependence compatible with the HERA measurements, one effectively takes $\lqcd$ to be a fit parameter, and the small value obtained for it is expected to capture most important higher order effects. Its value is also correlated with the value chosen for the infrared regulator $m$ in the JIMWLK evolution~\cite{Mantysaari:2018zdd}. Indeed first fits to HERA structure function data at next-to-leading order do not require as small values for $\lqcd$~\cite{Beuf:2020dxl}. 
This parametrization gives $\as = 0.14$ or $0.16$ at the typical scale $r=1/M_{\jpsim }$ depending on the value used for $\lqcd$.

\begin{figure}
    \centering
    \includegraphics[width=\columnwidth]{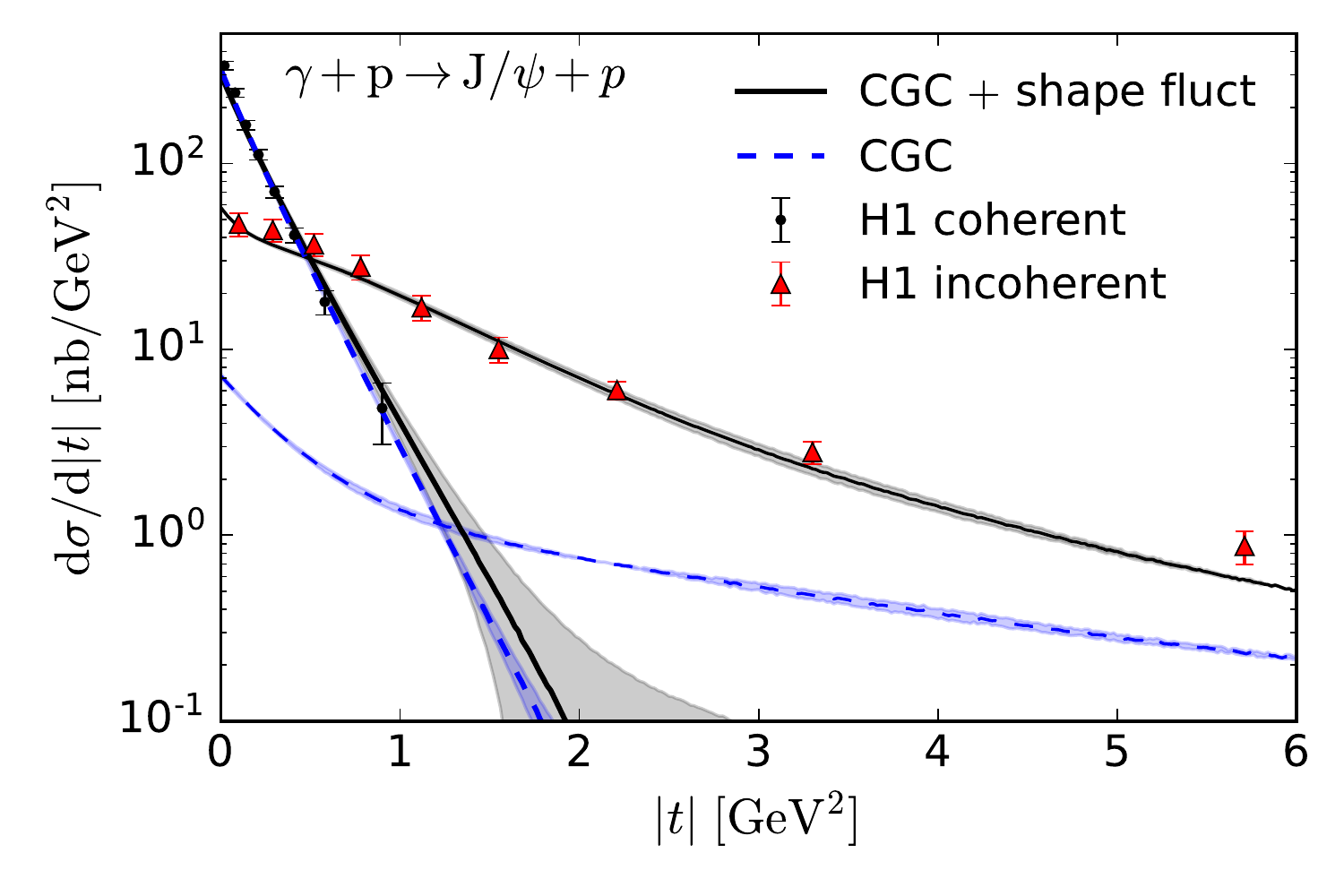}
    \caption{Coherent and incoherent \jpsi photoproduction cross section calculated from the CGC framework with and without proton shape fluctuations compared to the H1 data~\cite{H1:2013okq}. The bands show  statistical uncertainties of the calculations. }
    \label{fig:hera_coh_t_spectra}
\end{figure}

\begin{figure}
    \centering
    \includegraphics[width=\columnwidth]{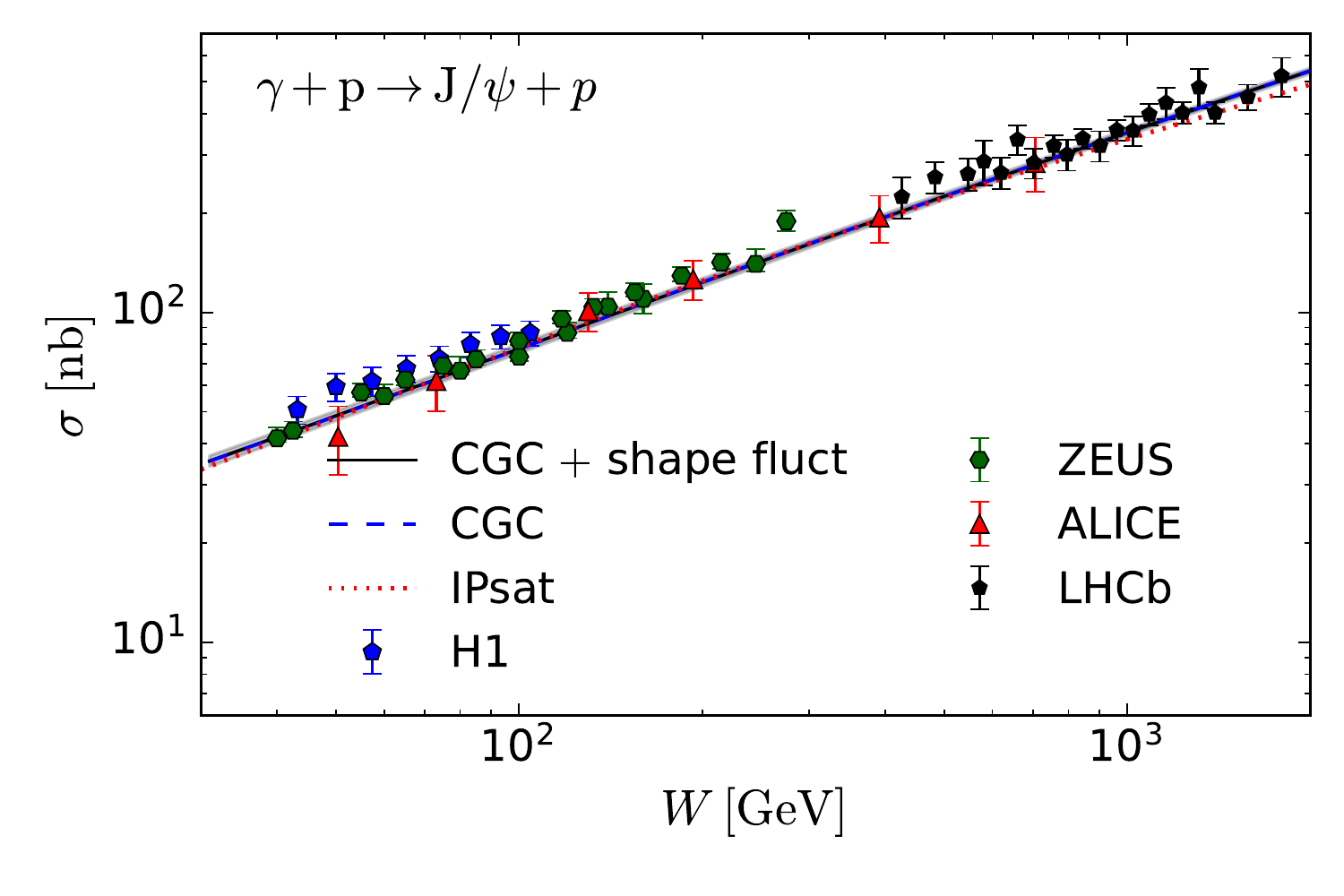}
    \caption{Total coherent \jpsi photoproduction cross section in $\gamma+p$ scattering compared to the H1~\cite{H1:2005dtp,H1:2013okq}, ZEUS~\cite{ZEUS:2002wfj}, ALICE~\cite{ALICE:2014eof,ALICE:2018oyo} and LHCb~\cite{LHCb:2014acg, LHCb:2018rcm} data. For comparison the calculation using the IPsat parametrization for the dipole amplitude from Ref.~\cite{Mantysaari:2018nng} is also shown.}
    \label{fig:jpsi_wdep}
\end{figure}

With these parameters, we obtain a good description of the \jpsi spectra at $W=75\,\gev$ as shown in Fig.~\ref{fig:hera_coh_t_spectra}.
The coherent spectrum constrains the size of the proton and the overall density, and the incoherent cross section determines the amount of fluctuations. In practice, the hot spot size determines the slope of the incoherent spectra in the $|t|\sim 1\ \gev^2$ region. The density ($Q_s$) fluctuations are most important at low $|t|$~\cite{Mantysaari:2016jaz}, and at high $|t|$ the color charge fluctuations that result in a powerlike incoherent spectrum become visible~\cite{Mantysaari:2019jhh} (see also Ref.~\cite{Kumar:2021zbn} where additional substructure at smaller distance scales is introduced, and Ref.~\cite{Demirci:2022wuy} for an analytical study of the role of the geometry and color charge fluctuations). Without shape fluctuations only the color charge fluctuations contribute to the incoherent cross section at all $|t|$.
Using the fitted parameters, an excellent agreement of the energy dependence over a wide center-of-mass energy range is also obtained as illustrated in Fig.~\ref{fig:jpsi_wdep}.

 Let us finally note that the framework applied here is applicable in the high energy limit where the parton densities in the nucleon are very large and the DGLAP scale evolution~\cite{Gribov:1972ri,Gribov:1972rt,Altarelli:1977zs,Dokshitzer:1977sg} can be neglected. We note that there are also other collinear factorization based approaches that can be used to describe exclusive vector meson production in ultra peripheral collisions using (generalized) parton distribution functions, see for example Refs.~\cite{Forshaw:1995ax,Deak:2020mlz,Guzey:2020ntc,Eskola:2022vpi}.


\section{Results}
\label{sec:results}
\subsection{Vector meson spectra at the LHC}
\label{sec:lhc_spectra}
\begin{figure*}
     \subfloat[Coherent cross section as a function of \jpsi transverse momentum squared.]{
         \includegraphics[width=0.48\textwidth]{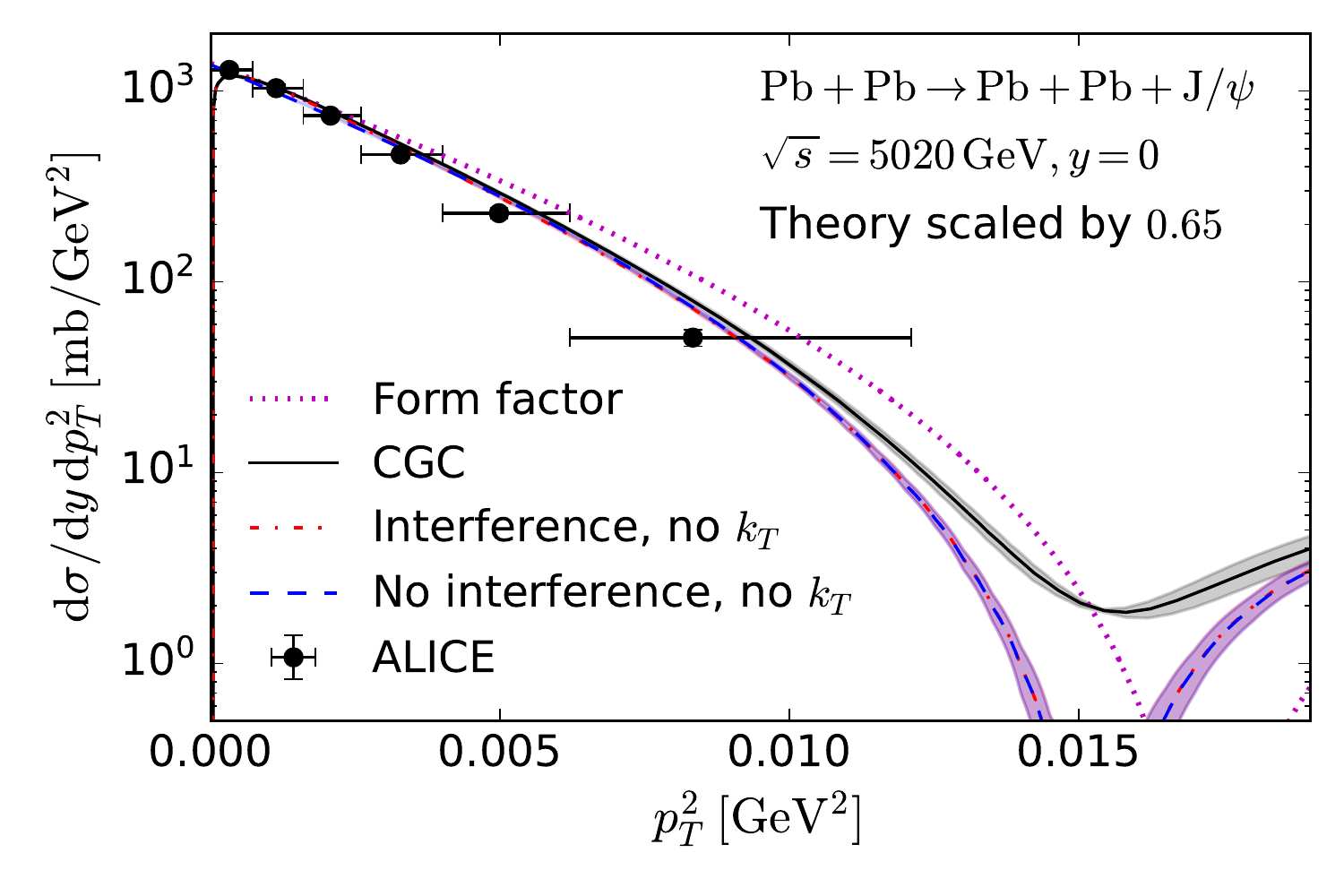}
         \label{fig:alice_spectra}
         }%
      \subfloat[Ratio to data in each experimental bin. Vertical error bars show the statistical uncertainty of the calculation, and the band shows the relative uncertainty of the ALICE data. ]{
         \includegraphics[width=0.48\textwidth]{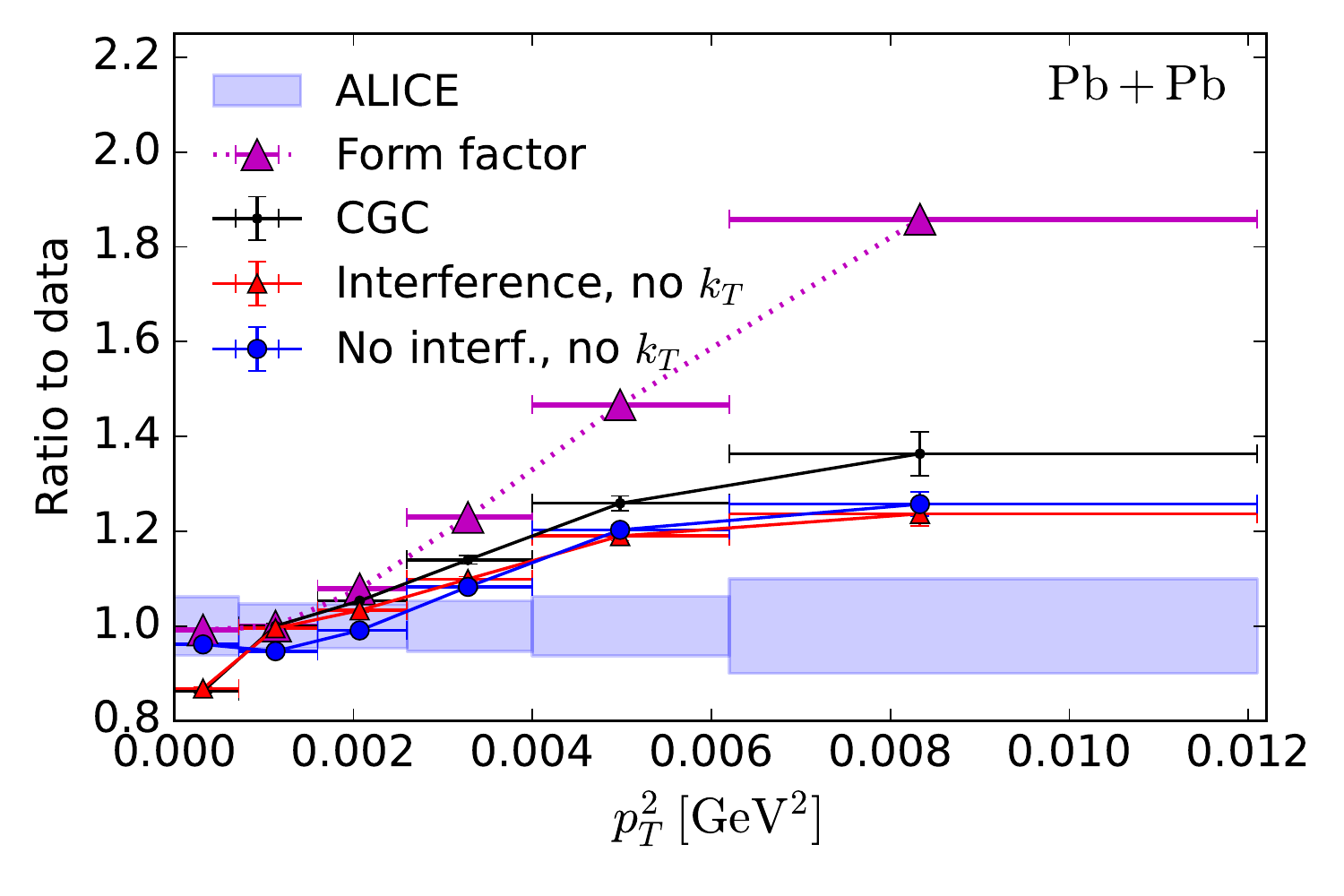}
         \label{fig:ratio_to_alice_w_interf}
         }%
     \caption{Coherent \jpsi\ production in ultra peripheral Pb$+$Pb collisions at $\sqrt{s}=5020\,\gev$. The theory calculations are normalized by a factor $0.65$, determined such that the full calculation matches the ALICE data in the second-to-lowest $\pt^2$ bin.  ALICE data~\cite{ALICE:2021tyx} (that corresponds to $\xpom\approx 0.0006$) includes statistical and uncorrelated systematical uncertainties. 
    }
     \label{fig:alice_with_interf}
\end{figure*}

The coherent \jpsi production cross section in ultra peripheral lead-lead collisions at the LHC as measured by the ALICE Collaboration~\cite{ALICE:2021tyx} and calculated from the CGC setup is shown in Fig.~\ref{fig:alice_with_interf}. 
The transverse momentum spectra are shown in Fig.~\ref{fig:alice_spectra}, and in order to more precisely compare the theory calculations to the experimental data
we show in Fig.~\ref{fig:ratio_to_alice_w_interf} the calculated cross sections divided by the ALICE data in the experimental \jpsi transverse momentum $p_T = |\pt|$ bins. 

The main result from our setup is labeled as \emph{CGC}, and includes saturation effects, a non-zero photon transverse momentum and the interference effect. We also show separately the result obtained by neglecting the photon transverse momentum $\kt$ but including the interference effect corresponding to Eq.~\eqref{eq:interference_only_midrapidity} (referred to as \emph{Interference, no $k_T$}), and by neglecting both the interference and the photon $\kt$ corresponding to Eq.~\eqref{eq:dsigma_dy_nointerf_nokt} (referred to as \emph{No interference, no $k_T$}). 
Nucleon substructure fluctuations are not included in any theory calculation here as they have a negligible effect on the shape of the coherent spectra. The dotted line (\emph{Form factor}) shows the squared two dimensional Fourier transform of the Woods-Saxon density profile, which is the result we would approximatively get in the absence of non-linear effects assuming that the dipole scattering amplitude is proportional to the nuclear thickness~\cite{Caldwell:2010zza} as e.g.~in the IPnonsat model discussed in Ref.~\cite{Mantysaari:2018nng}, and neglecting interference and the photon transverse momentum. 

The nucleon density is fixed in Sec.~\ref{sec:vm_production} by comparing to the HERA data, but the uncertainties in the data limit how accurately the proportionality constant between $Q_s$ and $g^2\mu$ in Eq.~\eqref{eq:qsmu}, which controls the overall normalization, can be determined.
However, as we will discuss in more detail below, this procedure in genreal leads to a too large normalization for the coherent cross section with nuclear targets compared to experimental data.
As at this point we are interested in the shape of the spectra, which probe the nuclear geometry, the theory calculations compared to the ALICE measurements are normalized by a constant factor determined such that the full CGC calculation matches the ALICE data in the second-to-lowest transverse momentum bin. Consequently, we only include statistical and uncorrelated systematical uncertainties (added in quadrature) to the experimental error bands that are shown in the figures. The applied normalization factor is shown in the figure.

\begin{figure}
    \includegraphics[width=\columnwidth]{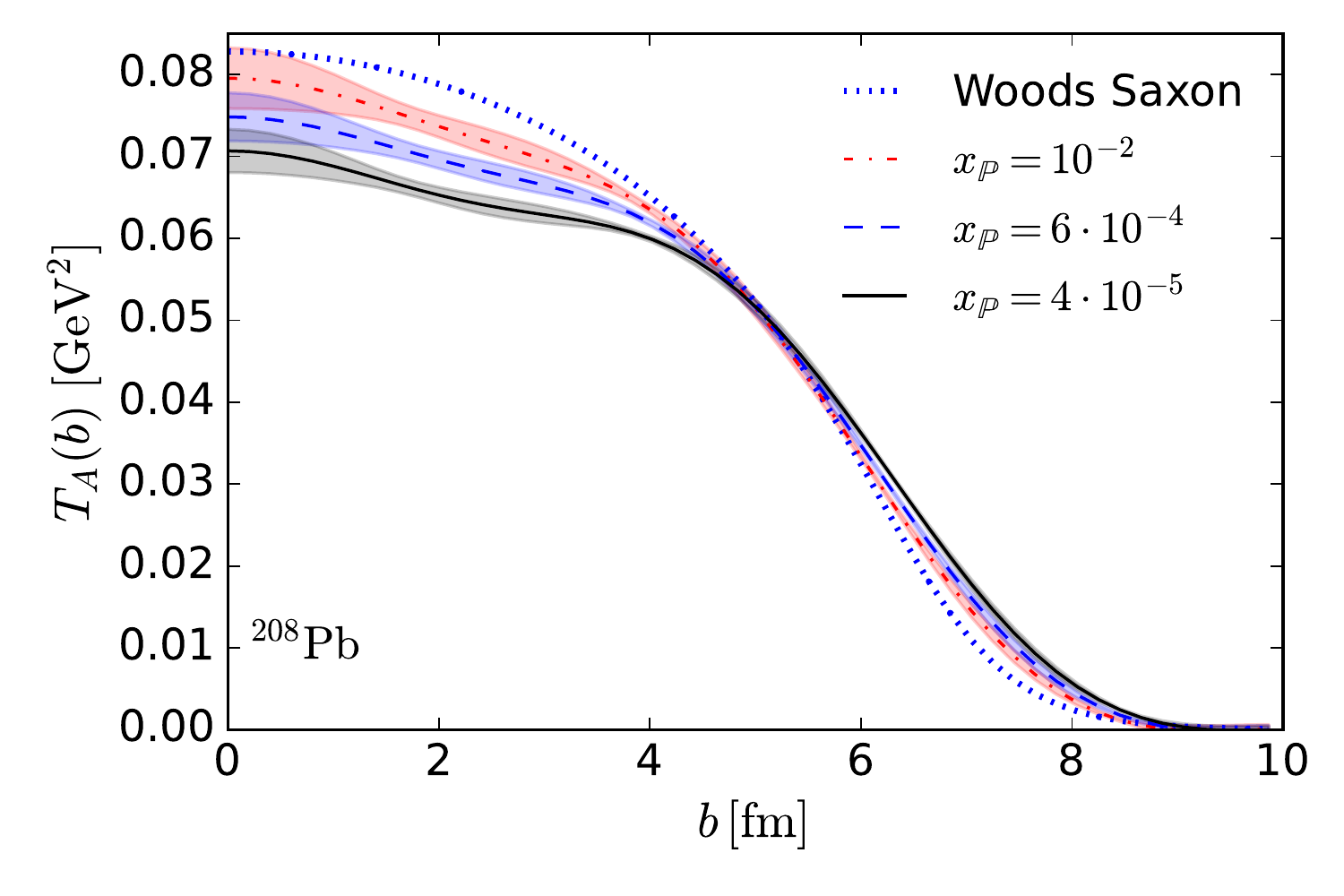}%
    \caption{Transverse density profile of the Pb nucleus at different $\xpom$ as extracted from the calculated $\gamma + \mathrm{Pb} \rightarrow \jpsim + \mathrm{Pb}$ spectra. The uncertainty band is obtained by varying the upper limit of the $|t|$ integration between $0.07 \dots 0.1\gev^2$ and the number of nuclear color charge configurations used to calculate coherent \jpsi production cross section. }
    \label{fig:ft_profile}
\end{figure}

The non-linear effects included in the CGC calculation are found to significantly improve the description of the ALICE data (we however note that a $p_T$ spectrum that differs from the form factor has also been obtained in Ref.~\cite{Guzey:2016qwo} without including non-linear dynamics). The fact that gluon saturation leads to a steeper spectrum is expected, as at the center of the nucleus one is closer to the black disc limit and the density profile of the nucleus starts to resemble that of a step function instead of the Woods-Saxon profile. We will demonstrate this effect in more detail later when discussing Fig.~\ref{fig:ft_profile}. However, even with the non-linear dynamics included we do not get as steeply falling spectra as seen in the ALICE data.
The photon transverse momentum has the important effect of smearing out the first diffractive minimum almost completely.

\begin{figure*}
    \subfloat[Low $p_T^2$ part of the coherent cross section as a function of squared momentum transfer.  ]{
         \includegraphics[width=0.48\textwidth]{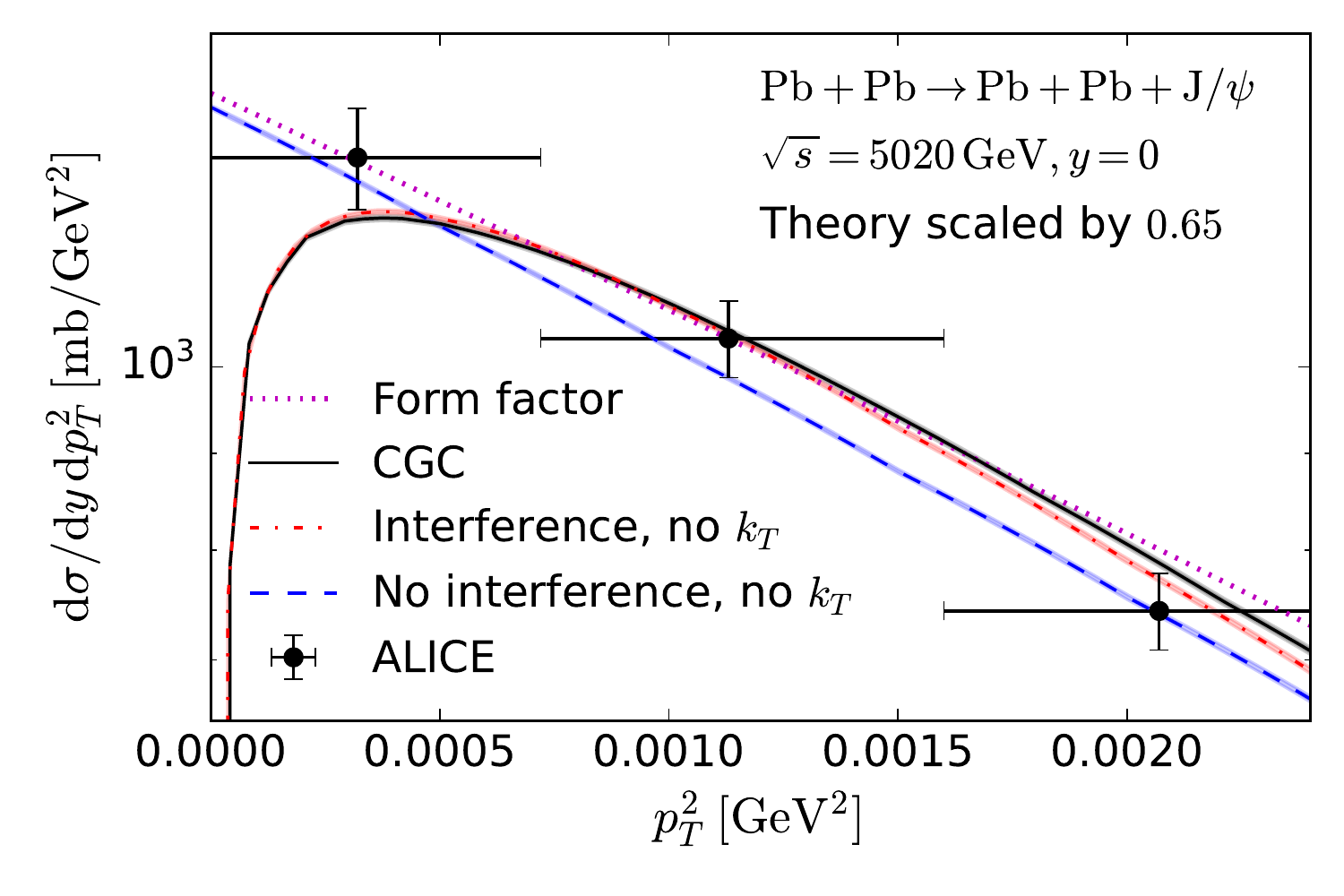}
         \label{fig:alice_pbpb_lowt}
         }%
\subfloat[Relative importance of the photon transverse momentum and interference effects]{
         \includegraphics[width=0.48\textwidth]{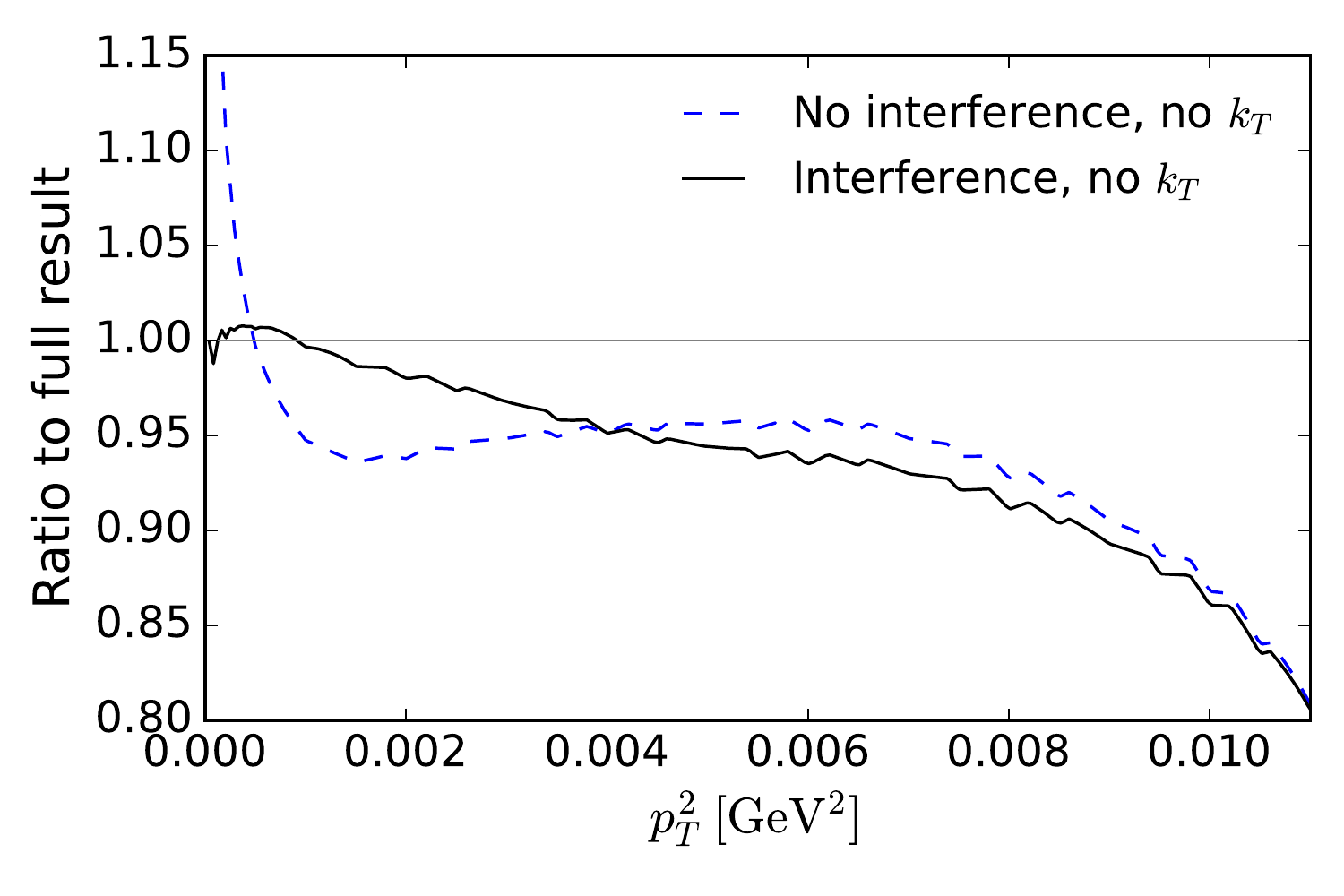}    
       
         \label{fig:midrapidity_effects_ratio}
     }
     \caption{Coherent \jpsi\ production in ultra peripheral Pb$+$Pb collisions at $\sqrt{s}=5020\,\gev$ compared to the ALICE data~\cite{ALICE:2021tyx} in the low \jpsi momentum region, and effect of the contributions from the interference effect and photon transverse momentum. }
     \label{fig:}
\end{figure*}

In order to illustrate in more detail the role of the interference effect we show the smallest $p_T^2$ part of the spectrum again in Fig.~\ref{fig:alice_pbpb_lowt}. Here we clearly see how the interference effect suppresses the cross section in the very low $p_T^2 \lesssim 0.0005\,\gev^2$ region. We note that the description of the ALICE data does not require the inclusion of this effect. 
To quantify the interference effect and the role of the photon transverse momentum in more detail we show in Fig.~\ref{fig:midrapidity_effects_ratio} the cross sections calculated by neglecting the photon transverse momentum, or both the photon $\kt$ and the interference effect, normalized by the full CGC result. Again the large interference effect at very small $p_T^2$ is clearly visible, as well as the fact that the interference effect becomes negligible  above $p_T^2 \gtrsim 0.005\,\gev^2$.

We note that although the interference effect is important especially at small $p_T^2$, and the photon transverse momentum significantly alters the spectra around the diffractive dip, these two effects in total increase the $p_T$ integrated cross section by only $\approx 3\%$. As the interference is maximal at midrapidity~\cite{Bertulani:2005ru}, we conclude that both of these effects have a negligible effect  on $p_T$ integrated cross sections that we study in Sec.~\ref{sec:tint_xs}.

\begin{figure*}
    \subfloat[Transverse momentum spectra ]{
         \includegraphics[width=0.48\textwidth]{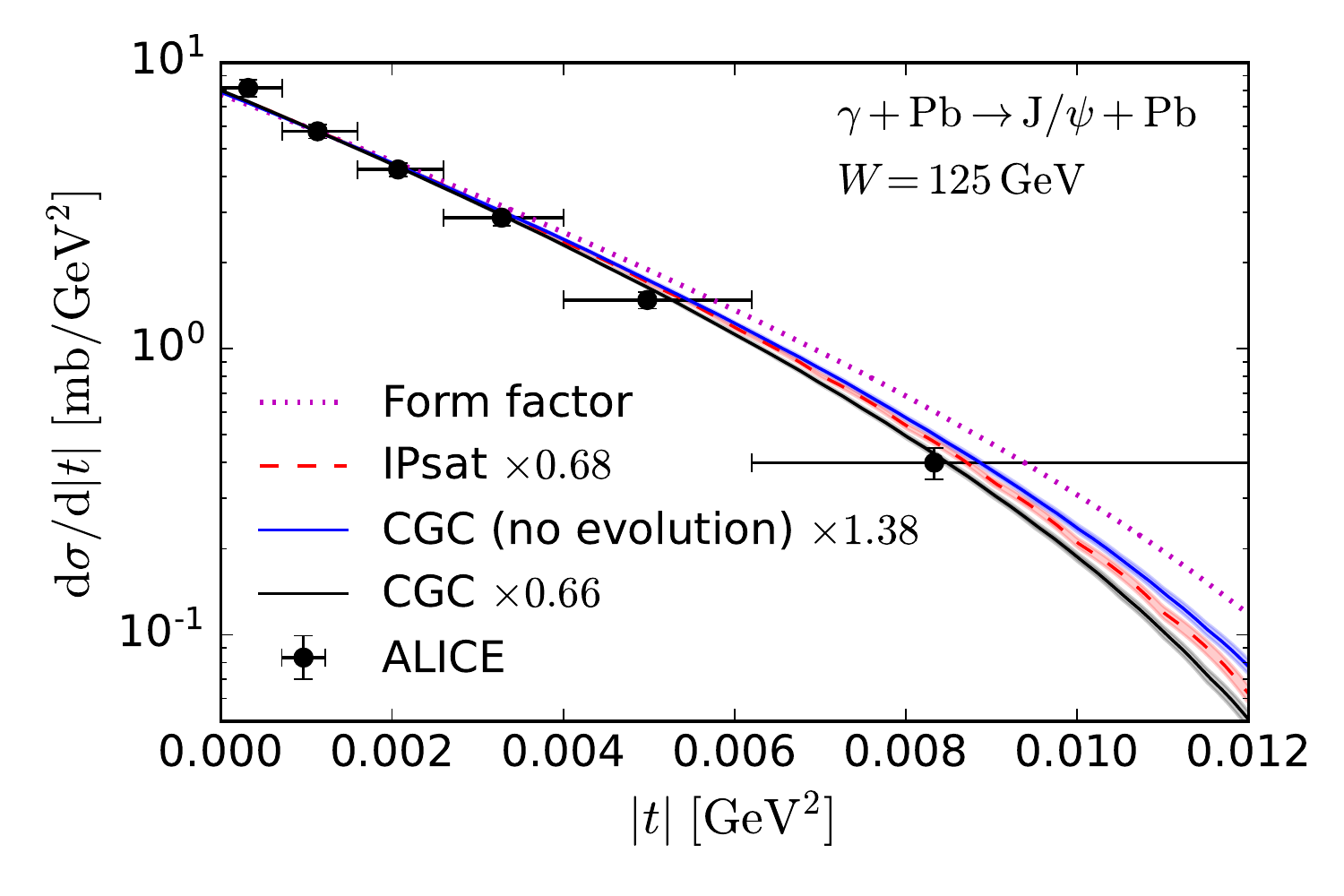}
         \label{fig:alice_spectra_nointerference}
         }%
\subfloat[Ratio to data in ALICE bins]{
         \includegraphics[width=0.48\textwidth]{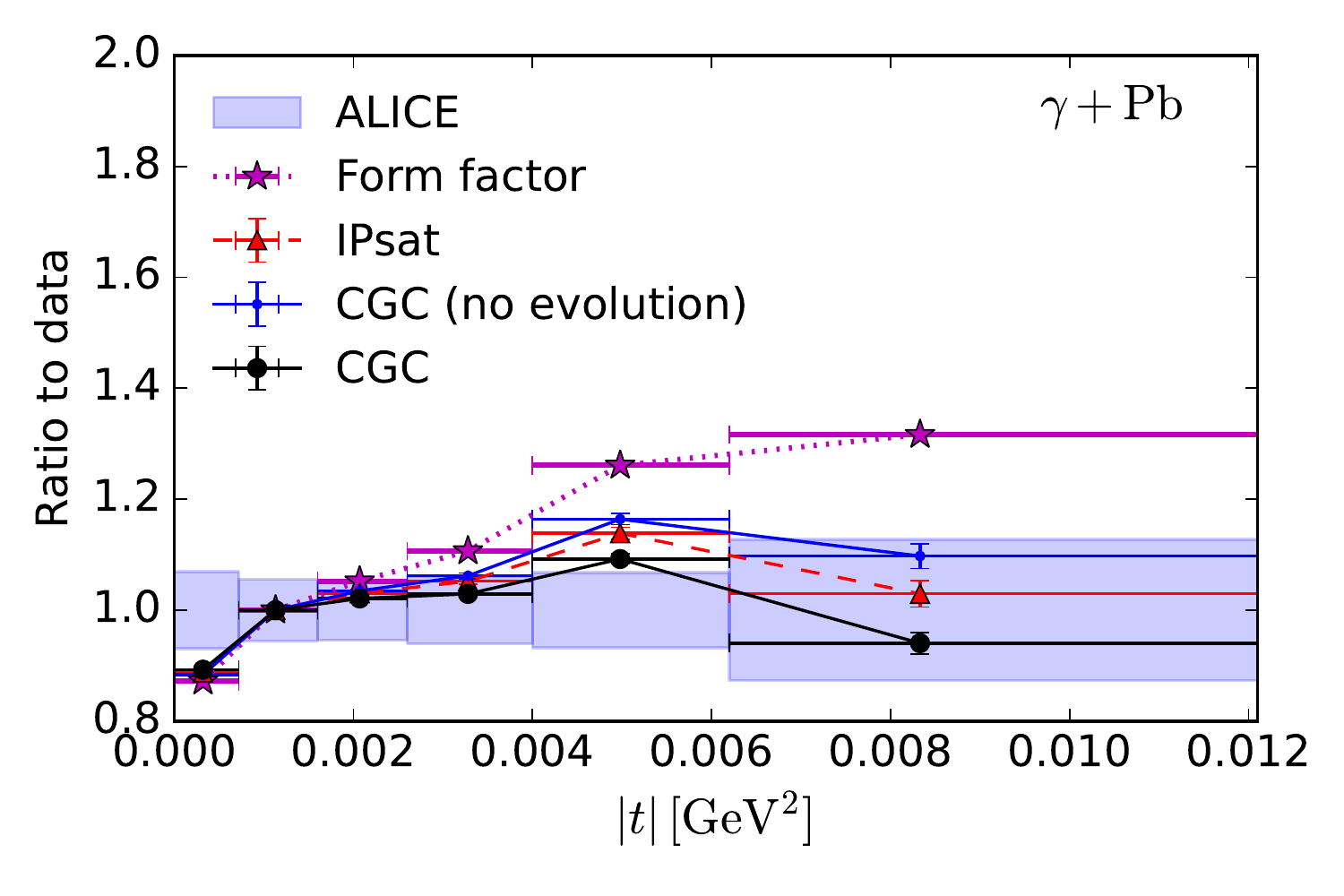}    
       
         \label{fig:alice_spectra_ratio_nointerference}
     }
     \caption{Coherent \jpsi\ production in $\gamma+\mathrm{Pb}$ collisions at $W=125\,\gev$ wich corresponds to midrapidity in ultra peripheral $\mathrm{Pb}+\mathrm{Pb}$ collisions at $\sqrt{s}=5020\,\gev$. The theory calculations are normalized to match the data in the second-to-lowest $t$ bin. The ALICE data~\cite{ALICE:2021tyx} includes statistical and uncorrelated systematical uncertainties. 
     }
     \label{fig:alice_gamma_Pb}
\end{figure*}

The ALICE Collaboration also reported in Ref.~\cite{ALICE:2021tyx} the differential cross section for the coherent \jpsi production in $\gamma + \mathrm{Pb}$ collisions, extracted from the measured cross section in Pb$+$Pb collisions. The data is reported as a function of squared momentum transfer $|t|$. In the frame where the photon has no transverse momentum $t \approx -\pt^2$. In practice, to extract the cross section for the  $\gamma + \mathrm{Pb}$ scattering the ALICE Collaboration removed the contribution from the photon transverse momentum and the interference effect. The ALICE data compared to the theoretical predictions calculated using Eq.~\eqref{eq:gamma_A_xs} are shown in Fig.~\ref{fig:alice_gamma_Pb} where we again show both the spectra and ratios to the data in the experimental squared momentum transfer bins. The theory calculations are normalized separately to describe the data at the second-to-lowest $|t|$ bin and the applied normalization factors are shown in the legends.

We again find that if non-linear dynamics is not included, the obtained \jpsi production spectra (Fourier transform of the form factor squared) is clearly less steep than the ALICE data. On the other hand, with non-linear dynamics included in the CGC setup a good description of the ALICE data is obtained except in the lowest $|t|$ bin where we underestimate the ALICE data. The disagreement in the smallest $|t|$ bin can be traced back to the fact that we predict a stronger interference effect in $\mathrm{Pb}+\mathrm{Pb}$ collisions than what is visible in the data, see Fig.~\ref{fig:alice_pbpb_lowt}.
We note that obtaining a steep enough spectrum in $\gamma+\mathrm{Pb}$ collisions from our setup, as we do, was not expected, as the shape of the Pb+Pb data is not well described (see Fig.\,\ref{fig:ratio_to_alice_w_interf}). 
In particular we note that in our calculation the non-zero photon transverse momentum renders the spectrum less steep when one moves from $\gamma+\mathrm{Pb}$ to $\mathrm{Pb}+\mathrm{Pb}$ collisions, which is opposite to what the ALICE data indicates.

In Fig.~\ref{fig:alice_gamma_Pb} we also show for comparison the result obtained by neglecting the JIMWLK evolution and evaluating the dipole amplitude at initial $\xpom=0.01$. As the evolution results in a larger nucleus, at smaller $\xpom$ the \jpsi spectrum is steeper and in a better agreement with the ALICE data.  We additionally show a result obtained using the IPsat parametrization from Ref.~\cite{Mantysaari:2018nng} (used e.g.~in Ref.~\cite{Mantysaari:2017dwh} to study \jpsi production in UPCs) for the dipole-nucleus scattering amplitude, in which case we obtain a very similar result as with the full CGC setup. A slightly less steep spectrum obtained with the IPsat dipole is expected, as the growing nuclear size with decreasing $\xpom$ is not included in the IPsat model.

\begin{figure}
    \includegraphics[width=\columnwidth]{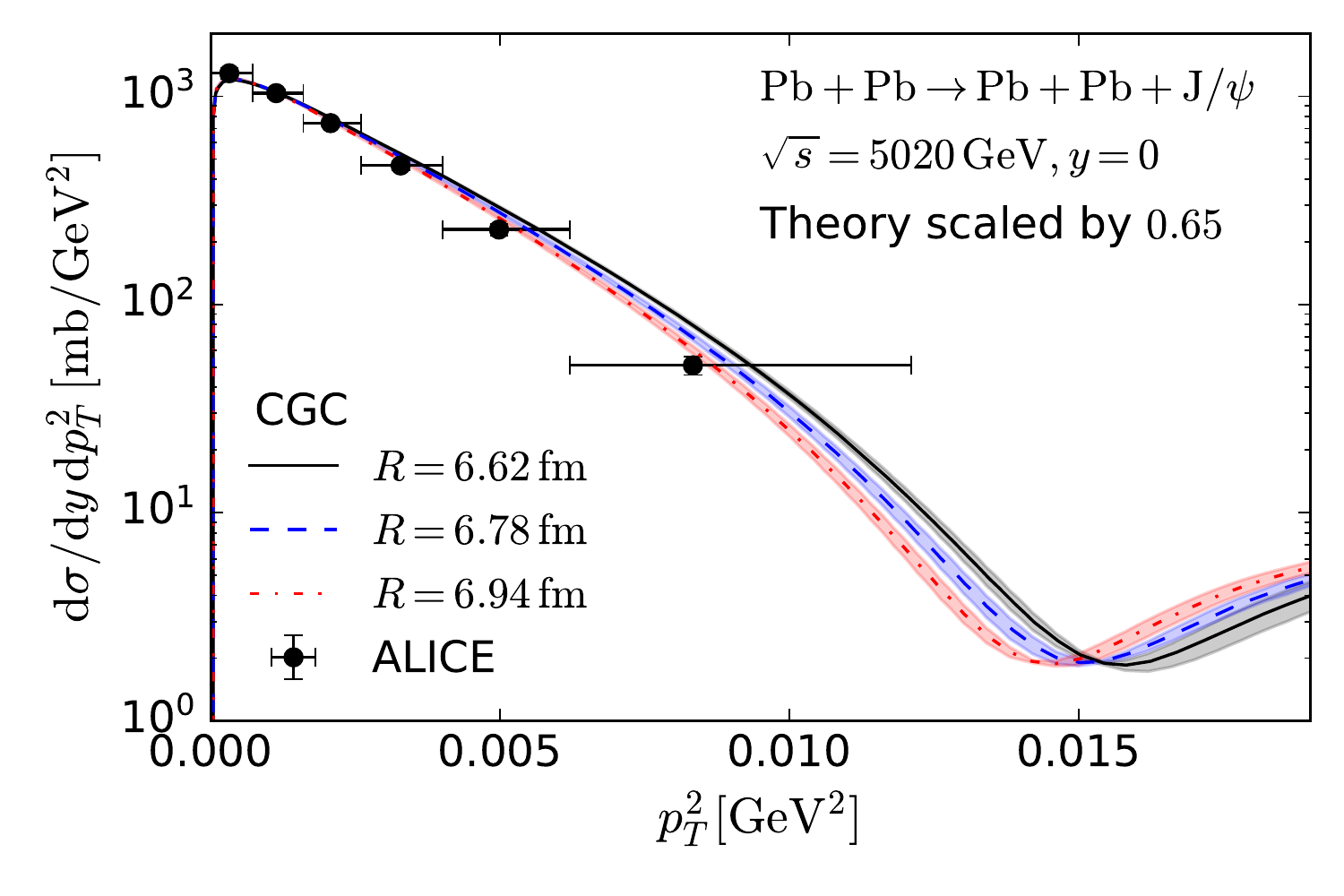}%
    \caption{Coherent \jpsi production cross section in Pb+Pb collisions calculated using the full CGC setup with different size parameters for the Pb nucleus at $\xpom=0.01$ compared to the ALICE data~\cite{ALICE:2021tyx}.}
    \label{fig:pbpb_size_dep}
\end{figure}

In the following we investigate the effect of different nuclear radius parameters in the Woods-Saxon parametrization. A larger nuclear radius is also motivated by the STAR Collaboration having measured exclusive $\pi^+\pi^-$ pair production in ultra peripheral Au+Au collisions, which indicates that the gluonic radius of the Au nucleus is larger than the nuclear charge radius extracted from low energy scattering experiments~\cite{STAR:2022wfe}. This is in contrast to the proton case, where the gluonic radius extracted from exclusive processes is smaller than the electromagnetic radius~\cite{Caldwell:2009ke}. Similarly, the analysis of Ref.~\cite{Albacete:2010sy} suggests that the heavy quarks in the proton, that originate from gluons, are located at a smaller transverse radius than the light quarks. 

We calculate coherent \jpsi production at midrapidity using three different radii for the Pb nucleus at $\xpom=0.01$ where the nucleon positions are sampled from the Woods-Saxon distribution. Our default choice is $R=6.62\,\mathrm{fm}$ for the radius parameter in the Woods-Saxon parametrization~\cite{DeVries:1987atn}. In Ref.~\cite{STAR:2022wfe} the STAR Collaboration extracts an approximately $4\%$ larger radius for Au compared to the standard literature value. The neutron skin effect can also result in an up to $\sim 0.2\,\mathrm{fm}$ larger gluonic size for the nucleus compared to the standard value determined from low-energy electromagnetic interactions~\cite{Tarbert:2013jze,PREX:2021umo}. Correspondingly we use 2.4\% and 4.8\% larger radii ($R=6.78\,\mathrm{fm}$ and $R=6.94\,\mathrm{fm}$) at the initial condition of the JIMWLK evolution. The resulting spectra in Pb+Pb collisions calculated from the full CGC setup where the interference effect and the photon transverse momenta are included are shown in Fig.~\ref{fig:pbpb_size_dep}.

Larger nuclei result in steeper spectra and an improved agreement with the ALICE $\mathrm{Pb}+\mathrm{Pb}\to \jpsim + \mathrm{Pb}+\mathrm{Pb}$ data. We note that we need a relatively large $R\approx 7\,\mathrm{fm}$ to describe the (gluonic) radius of the Pb nucleus at $\xpom=0.01$ in order to obtain a steep enough spectrum, compatible with the ALICE measurements. 
We emphasize that, as demonstrated above, the nonlinear dynamics also renders the spectra steeper by altering the spatial density profile of the nucleus as we will demonstrate explicitly below when discussing Fig.~\ref{fig:ft_profile}. Consequently without nonlinear dynamics an even larger nucleus would be needed in order to obtain a steep enough spectrum that describes the ALICE data.\footnote{Note that when using finite size nucleons, the Woods-Saxon parameters should be adjusted to recover the original distribution, but this leads only to an approximately 1\% increase in $R$ for the Pb nucleus \cite{Hirano:2009ah,Shou:2014eya}.}

The steeper spectra also result in smaller $p_T^2$ integrated cross sections. Compared to the default setup with $R=6.62\,\mathrm{fm}$, we obtain $2.5\%$ and $5.3\%$ smaller cross sections when using $R=6.78\,\mathrm{fm}$ or $R=6.94\,\mathrm{fm}$ nuclei. This is a non-negligible effect that should be kept in mind when also comparing the $p_T^2$ integrated cross sections to the LHC data. For the remainder of this paper, we will use the default value $R=6.62\,\mathrm{fm}$, which also results in a good agreement with the ALICE $\gamma + \mathrm{Pb} \rightarrow \jpsim + \mathrm{Pb}$ spectra, as shown in Fig.~\ref{fig:alice_gamma_Pb}.

\begin{figure*}
     \centering
    \subfloat[Full $p_T$ range ]{
         \includegraphics[width=0.48\textwidth]{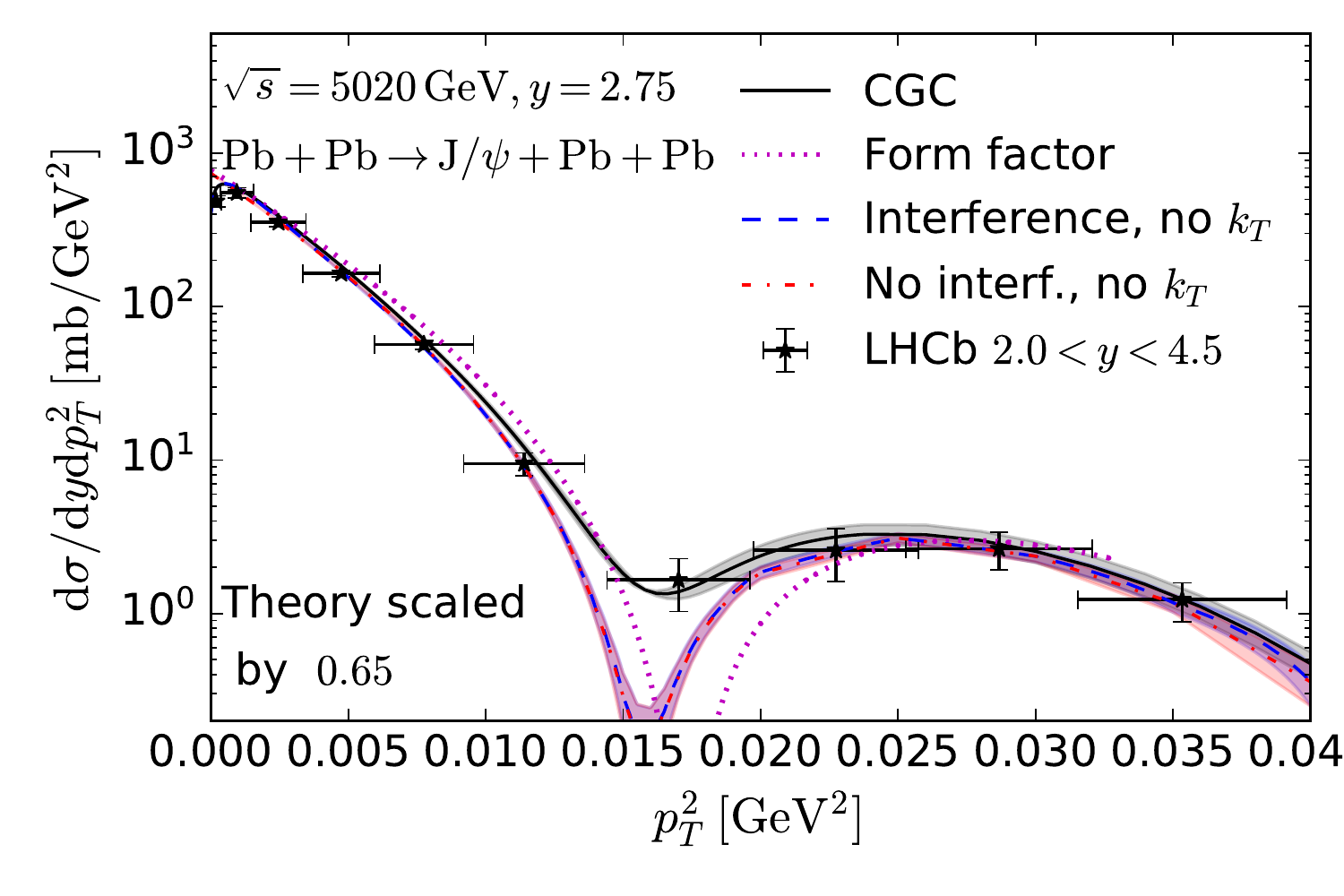}
         \label{fig:spectra_y_2.75}
         
        }%
     \subfloat[Low transverse momentum region ]{
          \includegraphics[width=0.48\textwidth]{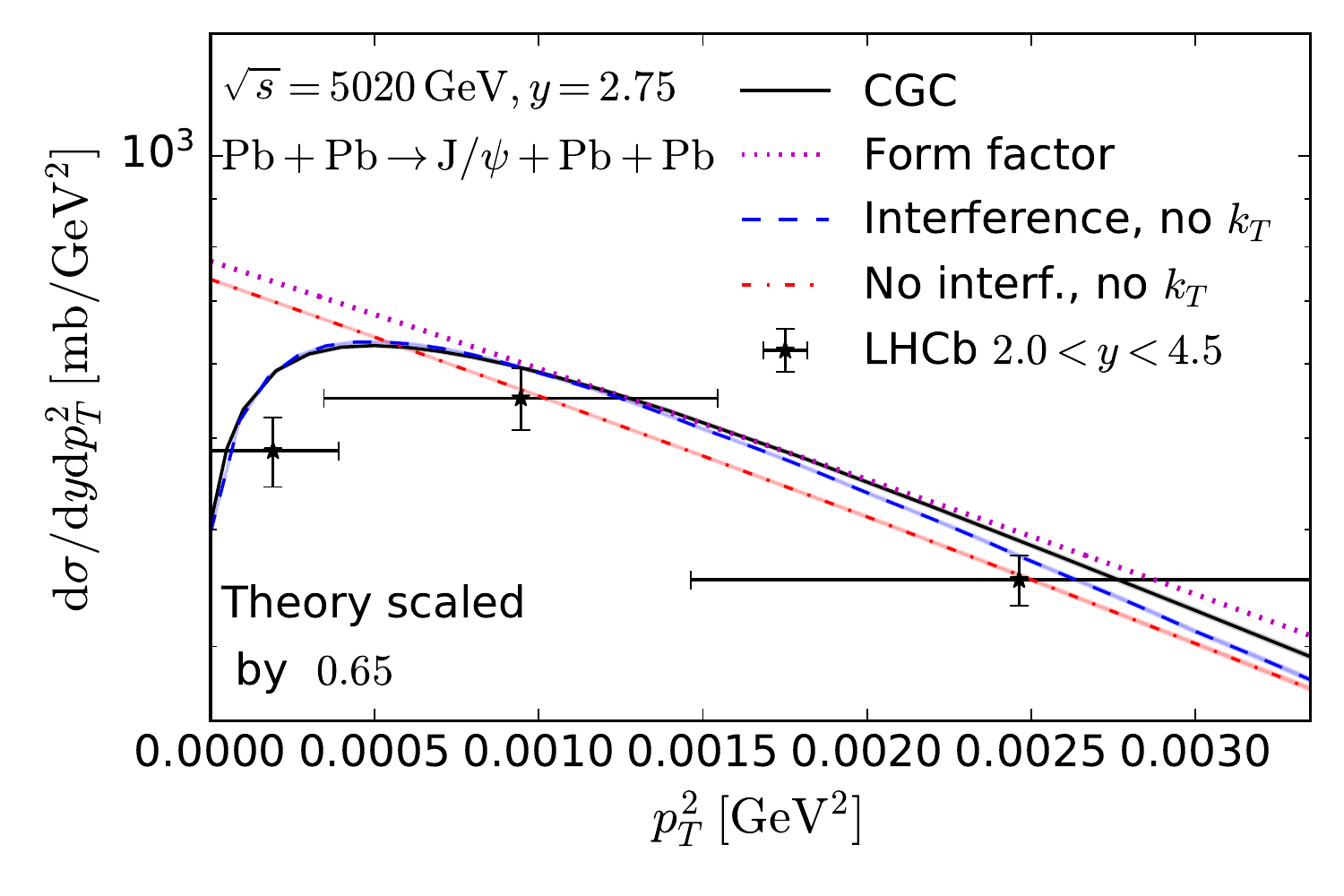}    
          \label{fig:spectra_y_2.75_smallt}
      }
      \caption{Coherent \jpsi production in ultra peripheral $\mathrm{Pb}+\mathrm{Pb}$ collisions at $\sqrt{s}=5020\,\gev$ at rapidity $y=2.75$ compared to the LHCb data~\cite{LHCb:2022ahs}. The theory calculations are normalized by the same factor as used when comparing to the ALICE midrapidity spectra.
      } 
      \label{fig:forward_spectra}
\end{figure*}

Next we study \jpsi production away from midrapidity where the first \jpsi transverse momentum spectrum has been measured recently~\cite{LHCb:2022ahs}.
In this case there are always  high-$\xpom$ and low-$\xpom$ contributions, 
and due to the different kinematics the two amplitudes for the $\gamma + \mathrm{Pb} \to \jpsim + \mathrm{Pb}$ scattering are not identical. Consequently the destructive interference at $p_T=0$ is not complete. This is demonstrated in Fig.~\ref{fig:forward_spectra} where we show predictions for the coherent \jpsi production in $\mathrm{Pb}+\mathrm{Pb}$ collisions at $\sqrt{s}=5020\,\gev$ in the forward region ($y=2.75$) as a function of squared \jpsi transverse momentum $p_T^2$. The cross sections are again calculated using the full setup, neglecting the photon transverse momentum, and neglecting both the photon $k_T$ and the interference effect. 

For comparison we show in Fig.~\ref{fig:forward_spectra} the LHCb data measured in the rapidity interval $2.0<y<4.5$. Note that we can only calculate 
scattering amplitudes below the initial value $\xpom=0.01$ used to initialize our setup, which limits our setup to be applicable only in the region $|y|< 2.79$ at $\sqrt{s}=5020\,\gev$. Consequently we can not calculate spectra in exactly the same kinematics as where the LHCb data is measured. 
Furthermore, the LHCb reports the cross section as $\mathrm{d}\sigma/\dd{y}y\dd{p_T}$ in transverse momentum bins. To transform this data to $\mathrm{d}\sigma/\dd{y}\dd{p_T^2}$ we calculate the mean transverse momentum $\langle p_T\rangle$ used in the Jacobian as
\begin{equation}
    \langle p_T \rangle = \frac{\int \dd{p_T} p_T \frac{\dd \sigma}{\dd{p_T} }} {\int \dd{p_T}  \frac{\dd \sigma}{\dd{p_T} }} \,,
\end{equation}
using the transverse momentum spectrum calculated from the full CGC setup that includes the interference effect and the photon transverse momentum. This transformation introduces an additional uncertainty not reflected in the experimental uncertainty bands especially in the smallest transverse momentum bins.

Keeping this difference in mind, we find a very good description of the LHCb spectra when both the interference effect and the photon transverse momentum are included. The overall normalization is again overestimated and consequently the theory curves in Fig.~\ref{fig:forward_spectra} are scaled by the same factor as above when compared to the midrapidity ALICE data. However as the rapidity bins in experimental data and theory calculation do not match, some differences in the absolute normalization are also expected.

Compared to the form factor, we again find that calculations including non-linear dynamics result in a steeper spectrum except in the dip region. However, compared to the midrapidity kinematics studied above, the difference between the full result and the Fourier transform of the form factor is smaller. This is due to the fact that a significant contribution to the cross section originates from the larger $\xpom \approx 0.01$ where non-linear effects are not as strong as at lower $\xpom$.

The fact that the destructive interference is not complete at zero $p_T$ is clearly visible in Fig.~\ref{fig:spectra_y_2.75_smallt}, where we see that at $p_T^2=0$ the interference effect suppresses the cross section only by roughly a factor $2$ in this kinematics. 
The interference effect is also clearly visible in the LHCb data, but note that there is some additional uncertainty in the Jacobian used to transform the experimental data to $\mathrm{d}\sigma/\dd{y}\dd{p_T^2}$. 
The photon transverse momentum again has a negligible effect at low $p_T^2$, and the calculation that includes the interference effect but no photon $k_T$ (Eq.~\eqref{eq:interference_only}) describes the spectra very accurately as long as one is far away from the diffractive minima.
 Around the first diffractive minimum at $p_T^2\approx 0.015\,\gev^2$ the non-zero photon transverse momentum again has a large effect and the first diffractive minimum is again almost completely removed.
 The interference effect and the photon transverse momentum in total have an $\approx 4.5\%$ effect on the $p_T$ integrated coherent cross section. 


Let us next illustrate the discussed role of saturation effects on the nuclear geometry. In Fig.~\ref{fig:ft_profile} we show the calculated transverse density profile of the nucleus, extracted form the calculated $\gamma + \mathrm{Pb} \rightarrow \jpsim + \mathrm{Pb}$ spectra (without nucleon substructure). The average transverse profile is obtained as a two dimensional  Fourier transform assuming that the scattering amplitude is purely real or purely imaginary (see e.g.~Refs.~\cite{STAR:2017enh,Klein:2019qfb}):
\begin{equation}
    T_A(b) \propto \int \Delta \dd{\Delta}  J_0(b \Delta) (-1)^n \sqrt{\frac{\der \sigma^{\gamma^* + \mathrm{Pb} \rightarrow \jpsim + \mathrm{Pb}}}{\der |t|}}, 
\end{equation}
and normalized such that $\int \dd[2]{b} T_A(b)=208$. The sign of the scattering amplitude changes at diffractive minima, which is taken into account by the factor $(-1)^n$ where $n$ is the number of minima at momentum transfers smaller than $\Delta$. For comparison the corresponding profile function calculated from the Woods-Saxon distribution is also shown. We note that the Woods-Saxon distribution is used as an input from which the locations for protons and neutrons are sampled at $\xpom=10^{-2}$. The $\xpom$ values shown in the figure correspond to midrapidity ($\xpom \approx 6 \cdot 10^{-4}$) and forward $y=2.75$ kinematics ($\xpom \approx 10^{-2}$ and $\xpom \approx 4 \cdot 10^{-5}$) at the LHC. 

The fact that nonlinear effects render the density profile flat at small impact parameters is clearly visible especially at smaller $\xpom$. In the dipole picture used in this work, this is a natural consequence of the unitarity bound for the dipole-nucleus scattering amplitude. At large impact parameters the growth of the nucleus toward small-$\xpom$ as a result of the JIMWLK evolution is also clearly visible.


\begin{figure}
    \includegraphics[width=\columnwidth]{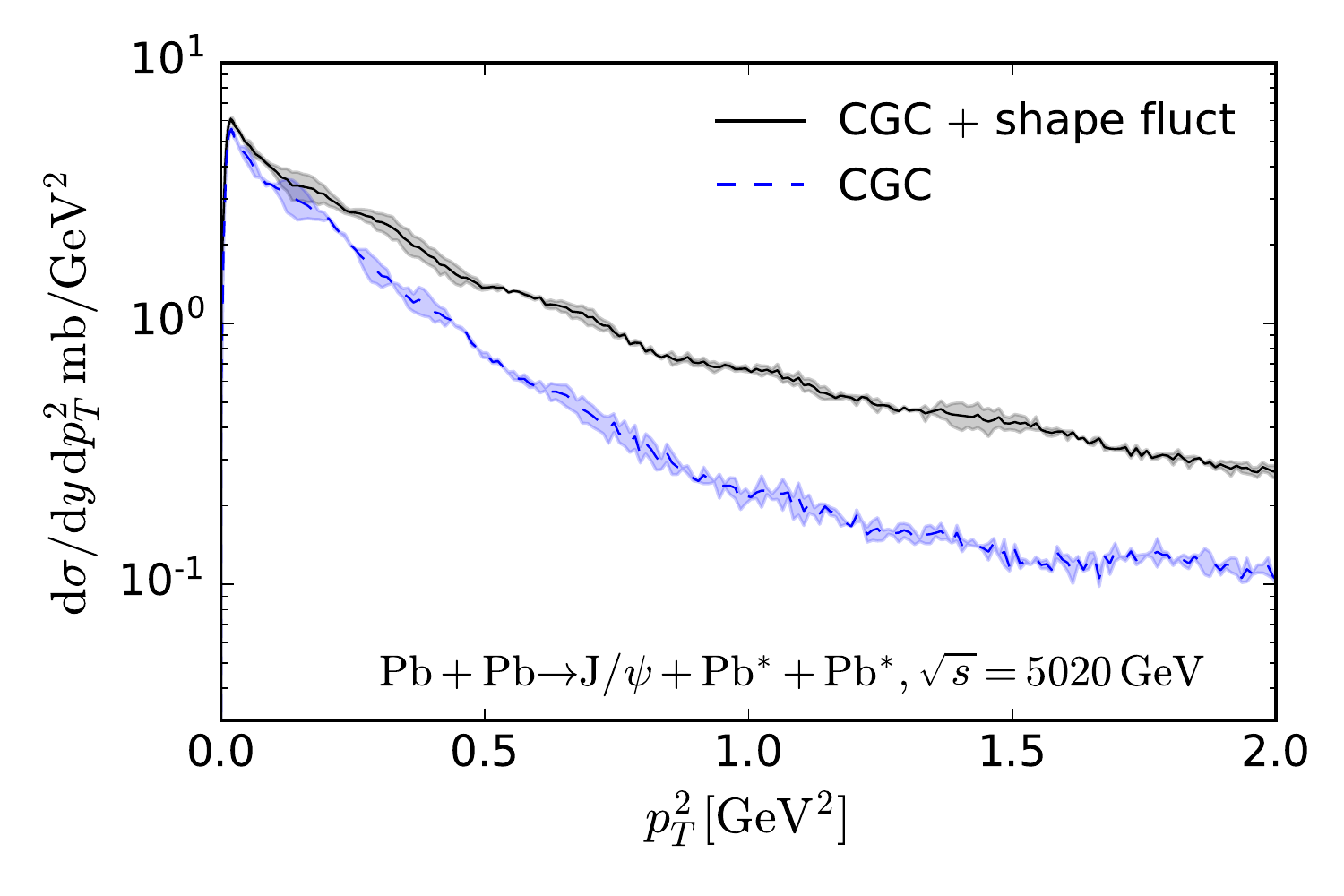}%
    \caption{Incoherent \jpsi production cross section as a function of squared meson transverse momentum $p_T^2$ in Pb$+$Pb collisions at $\sqrt{s}=5020\,\gev$ at midrapidity, calculated with and without nucleon shape fluctuations. The photon transverse momentum is neglected. 
    No artificial normalization factors are applied to the calculated spectra.
    }
    \label{fig:incoherent_5020_tspectra}
\end{figure}

At larger $p_T^2 \gtrsim 0.05\,\gev^2$ the incoherent production dominates. As the incoherent process is sensitive to the event-by-event fluctuations of the scattering amplitude (see discussion in Sec.~\ref{sec:vm_production} and Ref.~\cite{Mantysaari:2020axf} for a review), we next study exclusive \jpsi production as a function of $p_T^2$ focusing on the large momentum transfer region. In this kinematical domain the photon transverse momentum is negligible compared to the \jpsi transverse momentum (note that there are no dips in incoherent spectra around which the small photon $\kt$ would have a dominant effect) and we do not include it in the calculations. Furthermore the interference effect is also not visible in this kinematical domain as demonstrated in Appendix~\ref{appendix:interference}.

The incoherent cross sections calculated with and without nucleon substructure fluctuations are shown in Fig.~\ref{fig:incoherent_5020_tspectra}.
The substructure fluctuations are found to increase the incoherent cross section significantly by resulting in a less steeply falling spectrum at $p_T^2\gtrsim 0.25\,\gev^2$. At lower $p_T$ the incoherent cross sections are approximately identical. Physically, the substructure has an effect in the large $p_T^2\gtrsim 0.25\,\gev^2$ region, because at high momentum transfer one is sensitive to fluctuations at short distance scales. In fact, if the dipole scattering amplitude is proportional to the local density $T(\bt)$, then the width of the smallest fluctuating constituents determines the $p_T^2$ slope of the incoherent cross section at high $p_T^2$ as shown in Ref.~\cite{Lappi:2010dd} (neglecting the color charge fluctuations). On the other hand, at lower $p_T^2$ fluctuations at longer distance scale, the fluctuating positions of the protons and neutrons that are the same in both setups, dominate. 

These results are similar to what has been found using the IPsat parametrization for the dipole amplitude in Ref.~\cite{Mantysaari:2017dwh}. In addition to that previous analysis our framework also includes color charge fluctuations that e.g.~result in a non-zero incoherent cross section for the process $\gamma+p\to \jpsim + p^*$ even in the absence of geometry fluctuations, as demonstrated in Fig.~\ref{fig:hera_coh_t_spectra}. We find that the color charge fluctuations render the incoherent slope less steep at high $p_T^2 \gtrsim 0.6\,\gev^2$ in Pb$+$Pb collisions when nucleon substructure is not included, and the shape of the spectra changes from an exponential towards a power law as seen also in Fig.~\ref{fig:hera_coh_t_spectra} and in Ref.~\cite{Mantysaari:2019jhh}. When substructure fluctuations are included the effect of the color charge fluctuations is not as clearly visible in the studied $p_T^2$ range.


\subsection{Total coherent and incoherent cross sections at the LHC}
\label{sec:tint_xs}

To complete our discussion of the vector meson production in lead-lead collisions at the LHC, we show here the $p_T$ integrated \jpsi production cross sections as a function of \jpsi rapidity. 
As discussed previously, the interference effect and the photon transverse momentum have negligible effects on the integrated cross sections. We show results in the kinematics where $\xpom<0.01$.

\begin{figure*}
    \centering
    \begin{minipage}{0.48\textwidth}
    \includegraphics[width=\columnwidth]{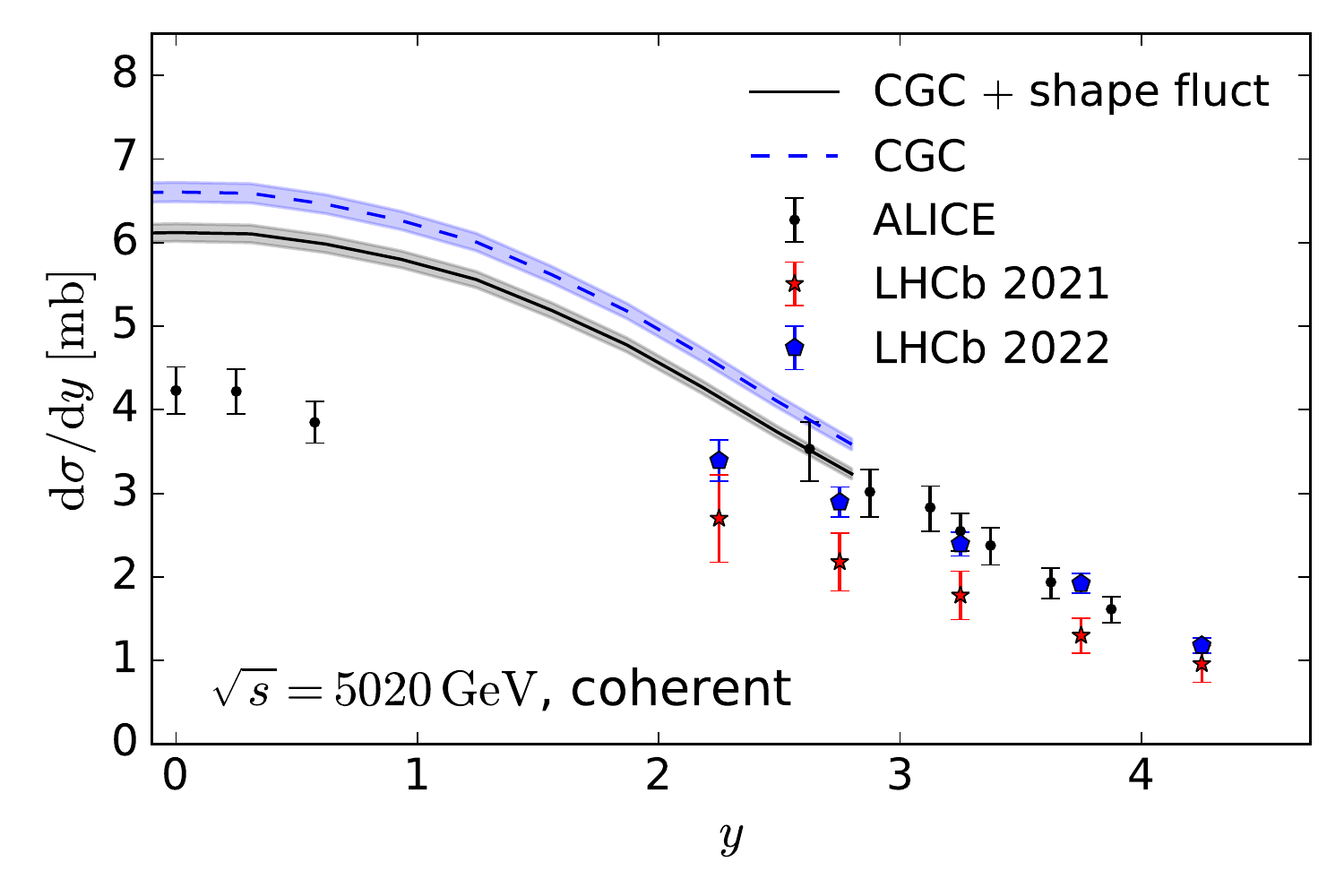}
    \caption{Coherent \jpsi production cross section  in ultra peripheral Pb$+$Pb collisions at $\sqrt{s}=5020\,\gev$ as a function of \jpsi rapidity compared to the ALICE~\cite{ALICE:2021gpt,ALICE:2019tqa} and LHCb 2021~\cite{LHCb:2021bfl} and 2022~\cite{LHCb:2022ahs} data.  }
    \label{fig:coherent_5020}
    \end{minipage}
    \hfill
 %
%
%
    \begin{minipage}{0.48\textwidth}
    \includegraphics[width=\columnwidth]{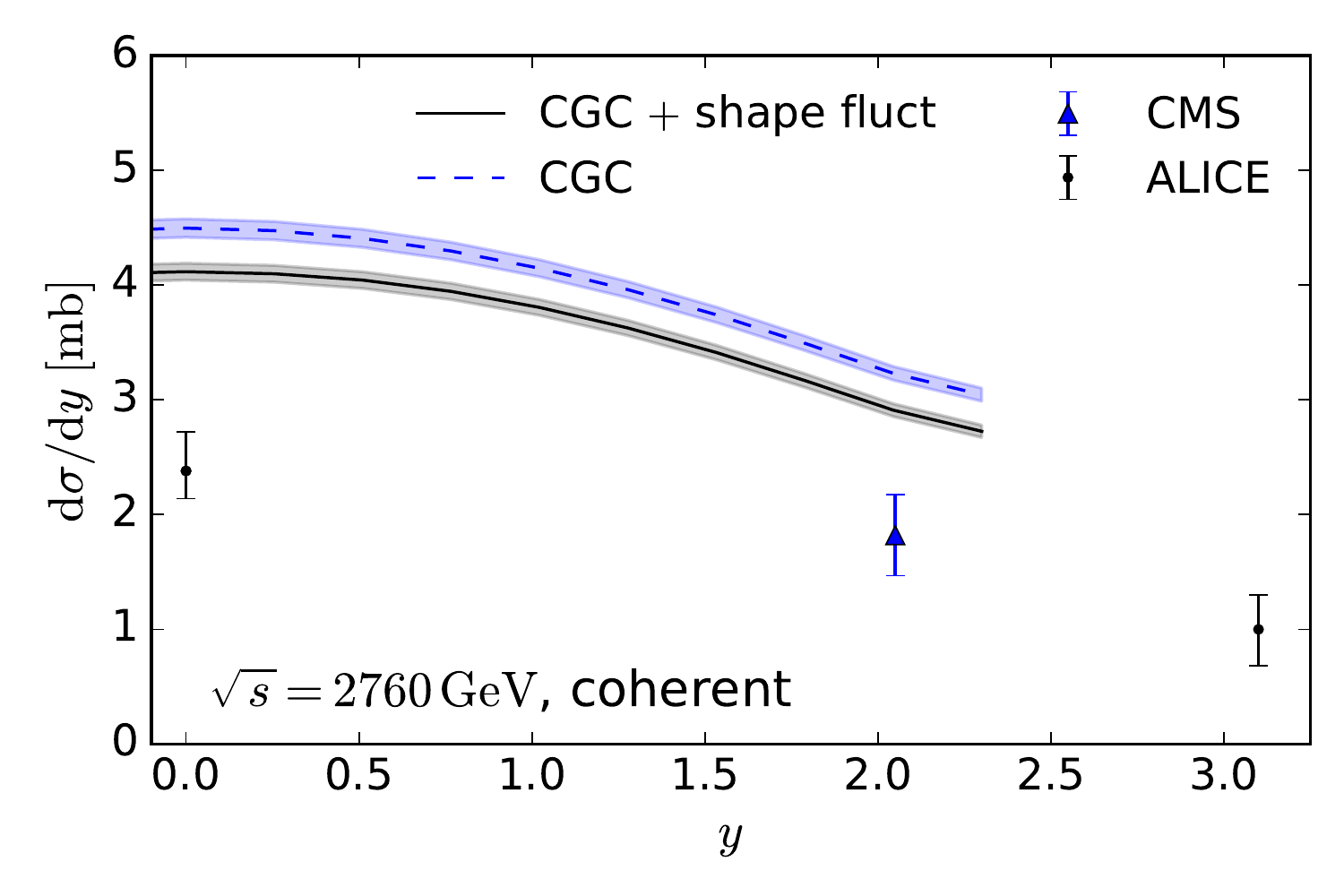}
    \caption{Coherent \jpsi production cross section  in ultra peripheral Pb$+$Pb collisions at $\sqrt{s}=2760\,\gev$ as a function of \jpsi rapidity compared to the ALICE~\cite{ALICE:2012yye,ALICE:2013wjo} and CMS~\cite{CMS:2016itn}  data.  }
    \label{fig:coherent_2760}
    \end{minipage}
\end{figure*}

We begin by comparing to the coherent cross section measurements at $\sqrt{s}=5020\,\gev$ by the ALICE~\cite{ALICE:2021gpt,ALICE:2019tqa} and LHCb~\cite{LHCb:2022ahs,LHCb:2021bfl} Collaborations. The results calculated with and without nucleon substructure fluctuations are shown in Fig.~\ref{fig:coherent_5020}.
Similarly, the results at $\sqrt{s}=2760\,\gev$ compared to the ALICE~\cite{ALICE:2012yye,ALICE:2013wjo} and CMS~\cite{CMS:2016itn} data are shown in Fig.~\ref{fig:coherent_2760}. Note that the extra normalization factors used above when comparing to the measured meson $p_T^2$  spectra  are not included here when we calculate the $p_T$ integrated cross sections in this work.

We find that both the CGC calculations (with and without nucleon substructure) describe the rapidity (or $\xpom = M_{\jpsim} e^{\mp y}/\sqrt{s}$) dependence of the experimental data reasonably well. 
The normalization is generically overestimated, except when comparing with the ALICE forward rapidity data.
We note that in our current setup the normalization is fixed by the HERA \jpsi production data in $\gamma+p$ collisions. Although the HERA data does not constrain the normalization precisely (see discussion in Sec.~\ref{sec:vm_production} and Fig.~\ref{fig:jpsi_wdep}), our results indicate that we find too small nuclear suppression compared to what the LHC data indicates. This appears to be a generic feature in dipole model calculations where nuclear suppression in exclusive \jpsi photoproduction is generically of the order of $\sim 30\%$~\cite{Lappi:2020ufv,Mantysaari:2020lhf} in the considered kinematics. This is  a smaller suppression than what the LHC $\mathrm{Pb}+\mathrm{Pb}$ data suggests when compared to the impulse approximation which corresponds to the scaled $\gamma+p$ cross section~\cite{CMS:2016itn} (see also Refs.~\cite{Guzey:2020ntc,Eskola:2022vpi} where the magnitude of the nuclear effects has been found to be compatible with the EPPS16~\cite{Eskola:2016oht} and nCTEQ15~\cite{Kovarik:2015cma} nuclear PDFs). Possible uncertainties in the \jpsi wave function can not completely explain the too small nuclear suppression found in our calculation~\cite{Lappi:2020ufv}. In Sec.~\ref{sec:lhc_spectra} we found that the shape of the $p_T^2$ spectra prefers a large nucleus (see Fig.\,\ref{fig:pbpb_size_dep}) which would reduce the $t$ integrated cross section by $\sim 6\%$. This reduction would result in a slightly better agreement with the LHC data.

The coherent cross section obtained with the nucleon shape fluctuations is $\sim 7\%$ smaller than what is obtained with spherical nucleons. We note again that the coherent cross sections in $\gamma+p$ collision are in practice identical as shown in Fig.~\ref{fig:jpsi_wdep} (difference is less than $1\%$ at $W=125\,\gev$ corresponding to midrapidity kinematics at $\sqrt{s}=5020\,\gev$)\footnote{Note that in Ref.~\cite{Mantysaari:2017dwh} the calculations with and without substructure fluctuations resulted in different coherent \jpsi photoproduction cross sections in $\gamma+p$ collisions.}. Although the coherent cross section is only sensitive to the average interaction with the target, we find larger nuclear suppression when substructure fluctuations are included. This is  because the hot spots can overlap in heavy nuclei, leading to very high local densities where the non-linear effects are stronger. 

\begin{figure*}
    \centering
    \begin{minipage}{0.48\textwidth}
    \includegraphics[width=\columnwidth]{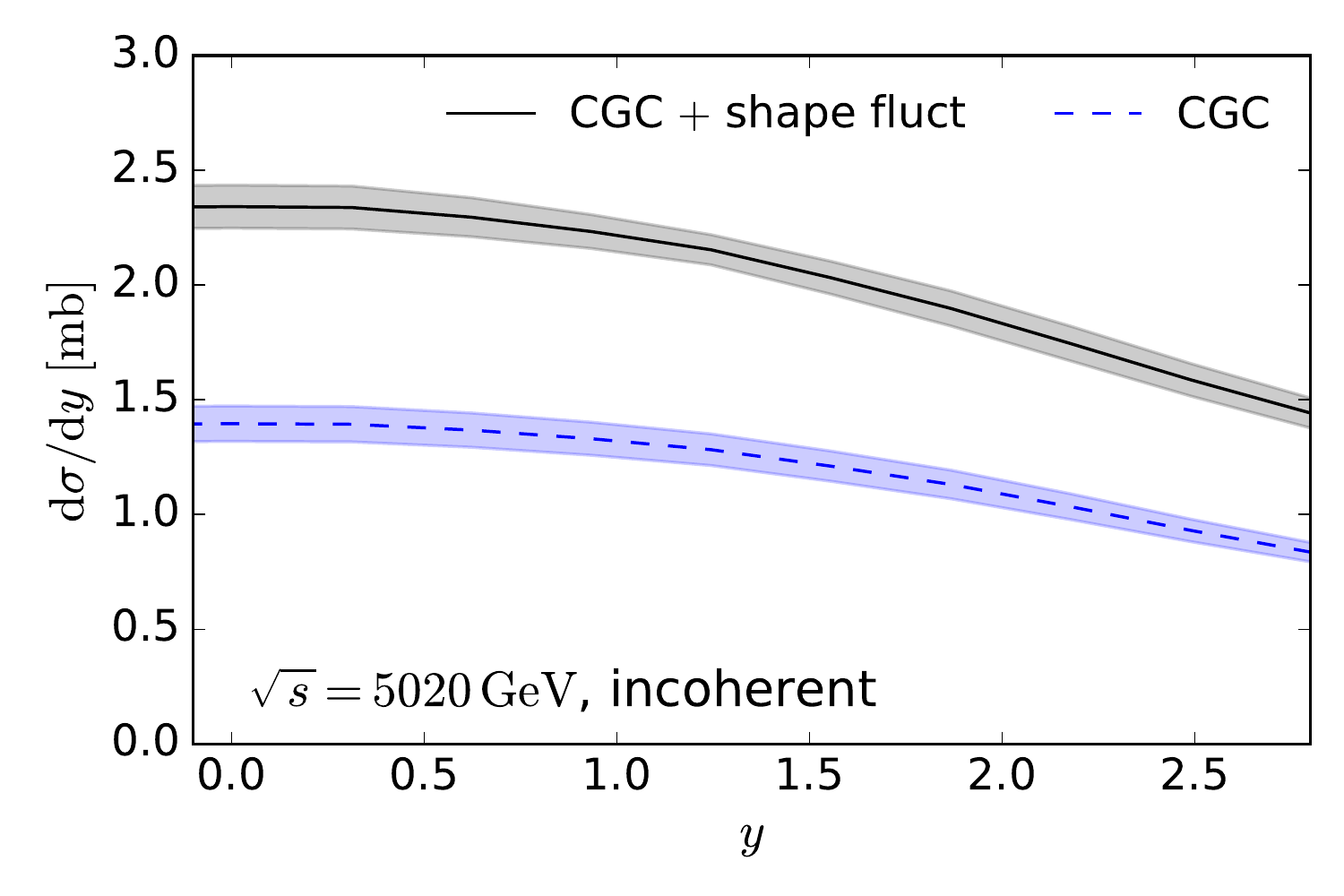}
    \caption{ Incoherent \jpsi production cross section  in ultra peripheral Pb$+$Pb collisions at $\sqrt{s}=5020\,\gev$ as a function of \jpsi rapidity.  }
    \label{fig:incoherent_5020}
    \end{minipage}
\hfill
    \centering
    \begin{minipage}{0.48\textwidth}
    \includegraphics[width=\columnwidth]{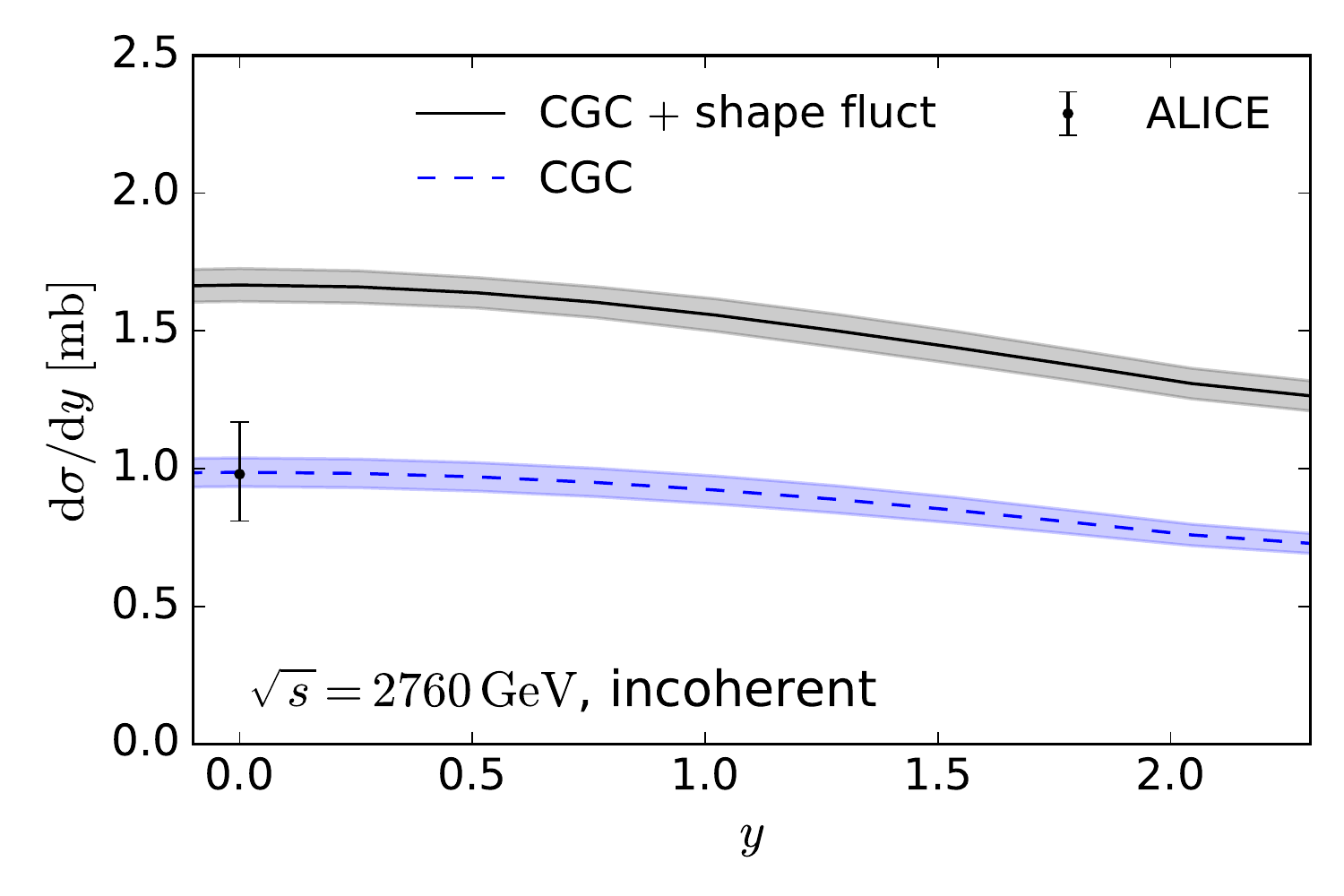}
    \caption{Incoherent \jpsi production cross section  in ultra peripheral Pb$+$Pb collisions at $\sqrt{s}=2760\,\gev$ as a function of \jpsi rapidity compared to the ALICE data\cite{ALICE:2013wjo}.  }
    \label{fig:incoherent_2760}
    \end{minipage}
\end{figure*}

The ALICE Collaboration has measured~\cite{ALICE:2013wjo} also the incoherent \jpsi production cross section at midrapidity at $\sqrt{s}=2760\,\gev$. 
Similar measurements can be expected at $\sqrt{s}=5020\,\gev$ in the near future. The incoherent cross sections calculated from the CGC setup with and without nucleon substructure fluctuations at both center-of-mass energies are shown in Figs.~\ref{fig:incoherent_5020} and~\ref{fig:incoherent_2760}.
We find that with the fluctuating substructure the measured incoherent cross section at $\sqrt{s}=2760\,\gev$ is  overestimated similarly to the coherent cross section studied above. The calculation with spherical nucleons describes the incoherent cross section data, but we note that it overestimates the coherent cross section significantly as seen in Fig.~\ref{fig:coherent_2760}. 

\begin{figure}
    \centering
    \includegraphics[width=\columnwidth]{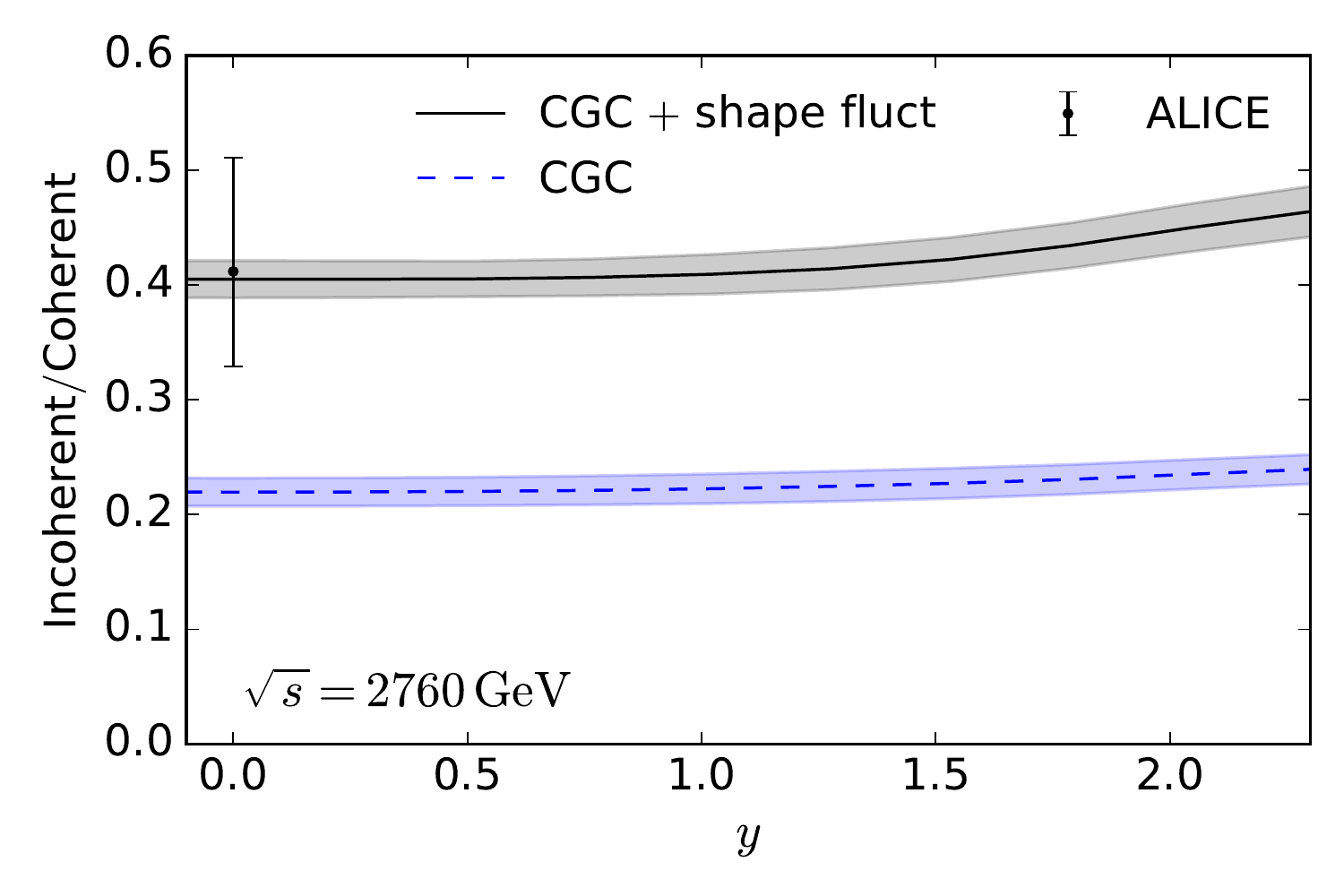}
    \caption{ Incoherent cross section divided by the coherent cross section as a function of \jpsi rapidity at $\sqrt{s}=2760\,\gev$. The ALICE data is calculated from cross sections reported in Ref.~\cite{ALICE:2013wjo}  by assuming completely uncorrelated uncertainties.  }
    \label{fig:ratio_2760}
\end{figure}
\begin{figure}
    \centering
    \includegraphics[width=\columnwidth]{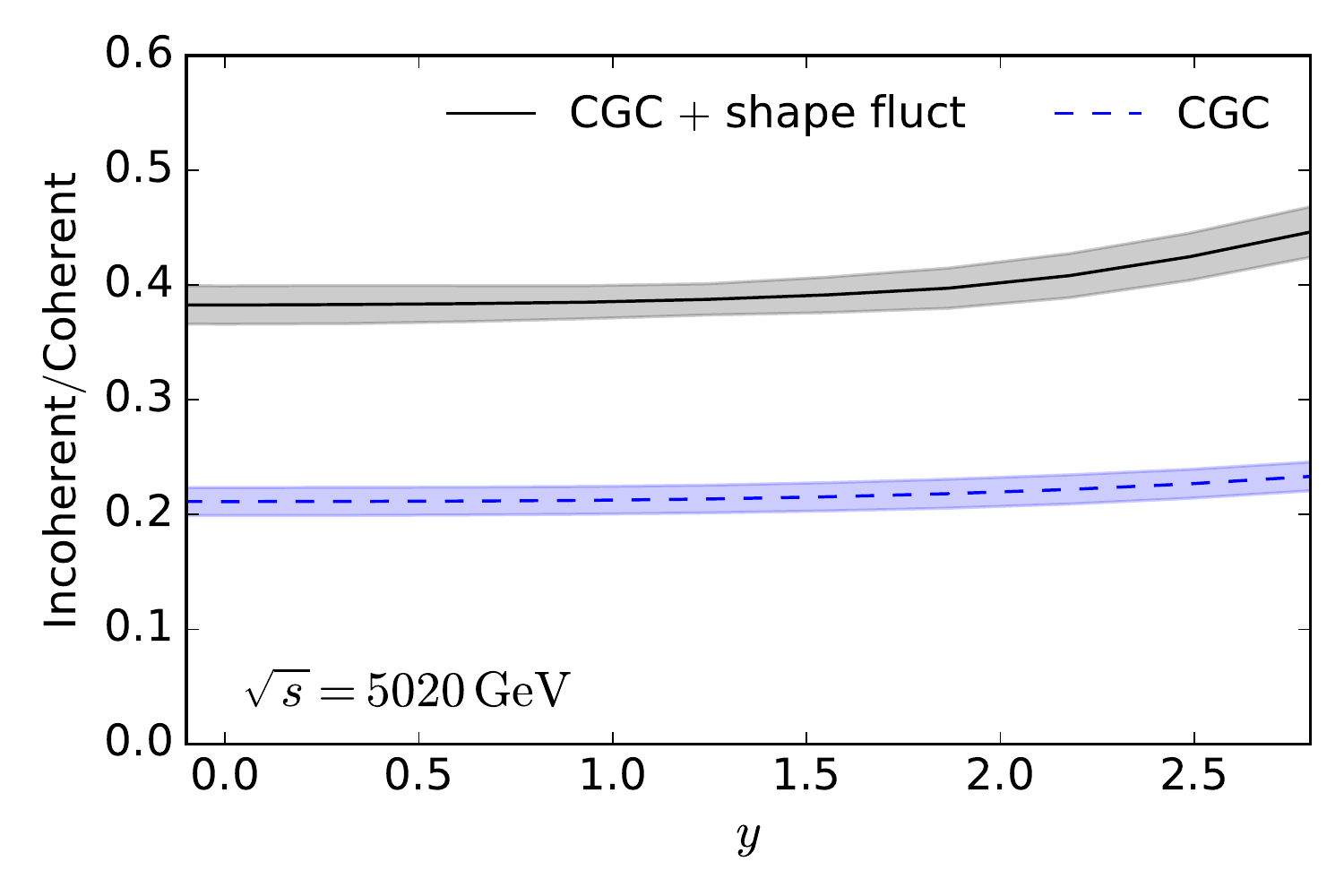}
    \caption{Predictions the ratio between the incoherent and coherent cross sections as a function of \jpsi rapidity at $\sqrt{s}=5020\,\gev$.  }
    \label{fig:ratio_5020}
\end{figure}

In order to cancel the normalization uncertainty we calculate the incoherent-to-coherent cross section ratio, and compare to the ALICE measurements. The results are shown in Fig.~\ref{fig:ratio_2760} at $\sqrt{s}=2760\,\gev$. The ALICE data point is calculated from the published coherent and incoherent cross sections~\cite{ALICE:2013wjo} assuming completely uncorrelated experimental uncertainties. Although our calculations overestimate both the coherent and incoherent cross sections, we find an excellent agreement with the experimental cross section ratio when nucleon substructure fluctuations are included in the calculation. The ratio is found to have a weak rapidity dependence originating from the geometry evolution which renders nucleons smoother at small $x$. This geometry evolution has only a moderate effect on the cross section ratio, as at large $|y|$ there is a two-fold ambiguity in the kinematics and one has to include both the small-$\xpom$ and large-$\xpom$ contributions. We note that in the IPsat model calculations in Ref.~\cite{Mantysaari:2017dwh}, results  with substructure fluctuations were also preferred by the ALICE data, but the cross section ratio was still slightly overestimated. A weaker rapidity dependence was also predicted from the IPsat setup as there is no geometry evolution. Predictions for the same ratio at $\sqrt{s}=5020\,\gev$ are shown in Fig.~\ref{fig:ratio_5020}, where a slightly smaller ratio is predicted. This is because at smaller $\xpom$ the nucleus is smoother, which suppresses event-by-event fluctuations and reduces the incoherent cross section.

\subsection{RHIC}
\label{sec:rhic}
In addition to the presented measurements from LHC, the STAR Collaboration at RHIC has measured exclusive \jpsi production in ultra peripheral collisions as a function of squared momentum transfer~\cite{star} at midrapidity at $\sqrt{s}=200\,\gev$. This measurement is sensitive to the nuclear structure at $\xpom = 0.0155$, and allows us to also study center-of-mass energy dependence. The initial condition for the JIMWLK evolution in this work is constrained at $\xpom=0.01$ (see discussion in Sec.~\ref{sec:vm_production}), and as such our framework is applicable only in the $\xpom \le 0.01$ region. 
Consequently, when comparing to the RHIC data we evaluate the dipole amplitude at $\xpom=0.01$ with the expectation that the small-$x$ evolution in this range has only a small effect on the cross section. However, we would generically expect this mismatch to result in too large overall normalization for the cross sections.

Let us first study coherent vector meson production. The STAR Collaboration has actually measured the total exclusive \jpsi production cross section~\cite{star}, which also includes an incoherent contribution. In order to extract coherent spectra, STAR subtracts the incoherent cross section calculated using a STARLIGHT~\cite{Klein:2016yzr} template, fitted to the data. As the incoherent cross section dominates for $p_T^2\gtrsim 0.02\,\gev^2$~\cite{Mantysaari:2017dwh}, comparisons to this preliminary data in the high $p_T^2$ region may not be robust.

\begin{figure*}
\subfloat[Full $p_T^2$ range]{
         \includegraphics[width=0.5\textwidth]{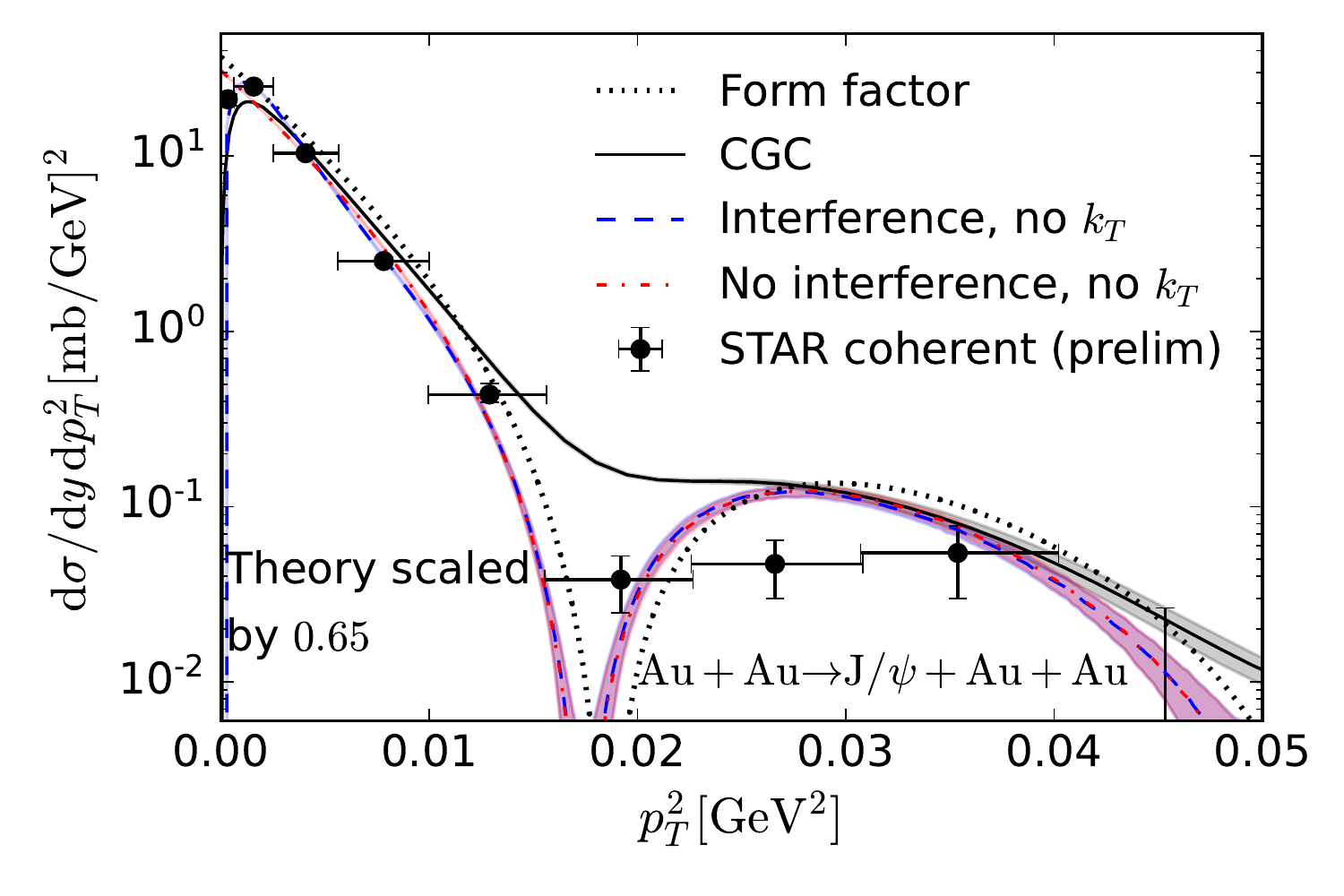}
         \label{fig:star_coherent_allt}
}%
\subfloat[Small $p_T^2$ region]{
         \includegraphics[width=0.5\textwidth]{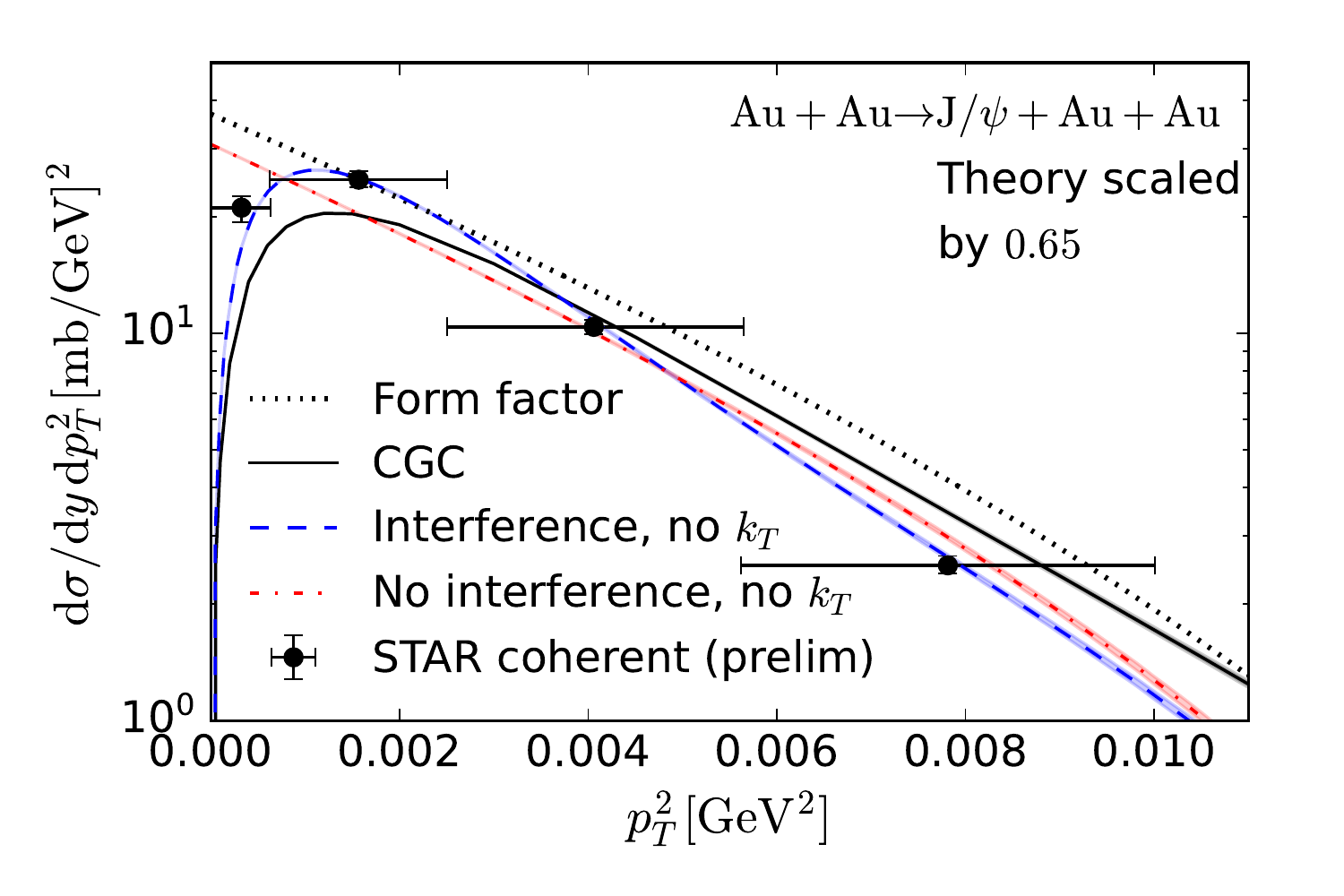}
         \label{fig:star_coherent_zoom}
}
     \caption{Coherent \jpsi production in ultra peripheral Au$+$Au collisions at $\sqrt{s}=200\,\gev$. Nucleon substructure fluctuations are not included in the calculation. The preliminary STAR data is extracted from Ref.~\cite{star} and only includes statistical uncertainties.  }
     \label{fig:star_coherent}
\end{figure*}

\begin{figure}

         \includegraphics[width=0.48\textwidth]{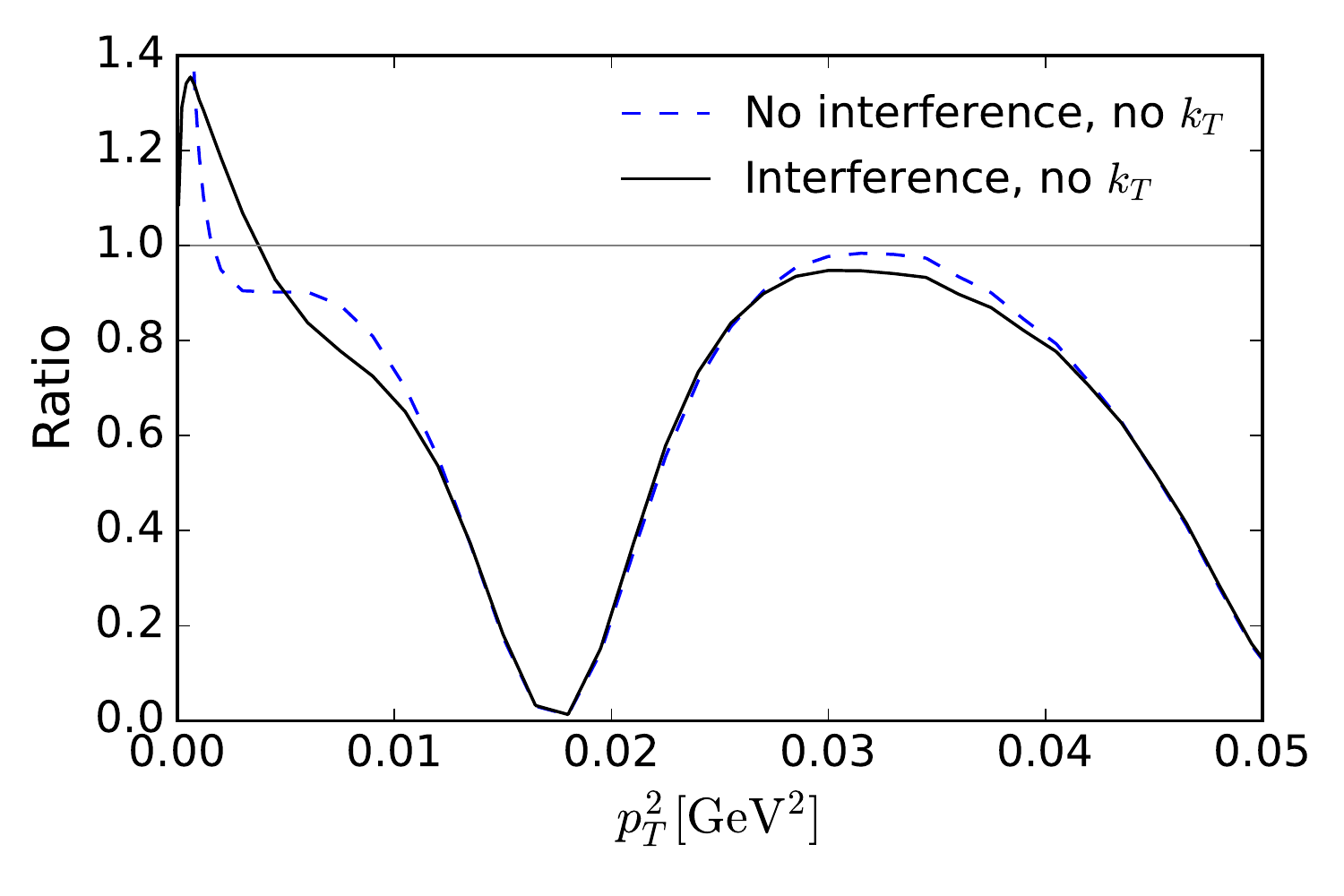}
         \label{fig:s}
         \caption{Differential coherent \jpsi production cross section in ultra peripheral collisions in RHIC kinematics as a function of squared \jpsi transverse momentum calculated with different assumptions normalized by the full result. 
        }
     \label{fig:rhic_interf_photonkt}
\end{figure}

The calculated coherent cross section is compared to the STAR data extracted from Ref.~\cite{star} in Fig.~\ref{fig:star_coherent}. The preliminary STAR data does not include systematic uncertainties, and consequently we only include STAR data in the $p_T^2\lesssim 0.05\,\gev^2$ region where the subtraction of the incoherent cross section can be expected to be most reliable. 
As in the LHC kinematics, we again calculate the differential cross section using our full setup, and neglecting the photon transverse momentum or both the interference effect and the photon transverse momentum.
The form factor is also shown in order to determine how the nonlinear effects change the spectra for RHIC kinematics. In order to quantify the importance of the  interference effect and the non-zero photon transverse momentum we show in Fig.~\ref{fig:rhic_interf_photonkt} the differential cross section calculated with different approximations normalized by the full result.

We again find that we predict a significantly larger normalization for the coherent cross section than seen in STAR data. Using the same normalization factor as was required to describe the ALICE transverse momentum spectrum a relatively good description of the overall normalization of the cross section is obtained. 
Unlike the ALICE measurement discussed in Sec.~\ref{sec:lhc_spectra}, the STAR data does not clearly distinguish between the calculations including saturation effects, and a simple Fourier transform of the Woods-Saxon density profile squared (\emph{Form factor} in Fig.~\ref{fig:star_coherent}). Both of these approaches result in comparable descriptions of the measured $p_T^2$ spectra.
Note that as shown in Fig.~\ref{fig:ft_profile}, the non-linear effects have only a moderate effect on the spatial structure of the nucleus in the $\xpom$ region probed at RHIC. When the photon transverse momentum is included, the cross section is significantly overestimated around the first diffractive minimum.

The interference contribution is found to be necessary to describe the fact that the cross section decreases towards $p_T \to 0$, visible in the smallest momentum transfer bin especially in Fig.~\ref{fig:star_coherent_zoom}. The photon transverse momentum has an even larger effect than in the LHC kinematics, and in particular the first diffractive minimum is now completely removed.
The stronger effect from the photon $\kt$ can be understood as the electromagnetic field of the Au nucleus dies much more rapidly as a function of distance at RHIC energies compared to the LHC (see Eq.~\eqref{eq:point_charge_field}), and smaller impact parameters correspond to larger photon transverse momenta.
The interference effect and the photon transverse momentum have an approximately $6\%$ effect on the $p_T^2$ integrated cross section. 
The importance of the interference effect as a function of meson transverse momentum is quantified in more detail in Appendix~\ref{appendix:interference}.

The PHENIX Collaboration has  measured~\cite{PHENIX:2009xtn} the total diffractive (sum of coherent and incoherent, although with the available statistics only a few events with $p_T^2$ in the region dominated by the incoherent cross section were observed) \jpsi production cross section at midrapidity in Au+Au collisions at $\sqrt{s}=200\,\gev$. The measured cross section is $\der \sigma^\text{Xn}/\der y= 76 \pm 35\,\mu$b for the process where at least one forward neutron is detected in zero degree calorimeters, and this requirement of the additional neutron emission is estimated to reduce the cross section by 45\%.  
Note that forward neutron emission is possible in coherent interactions, as a subsequent photon exchange can excite the nucleus. 
We get $\der \sigma^\text{coh}/\der y = 159\,\mu$b for the coherent cross section at midrapidity from the full setup, and taking into account the approximate reduction by 45\% when comparing to the PHENIX data with forward neutron emission, this corresponds to $\der \sigma^\text{Xn, coh}/\der y = 87 \, \mu$b. This is compatible with the PHENIX data especially when noting that we evaluate the dipole amplitude at smaller $\xpom$ than what the RHIC kinematics actually corresponds to.

\begin{figure}
    \centering
         \includegraphics[width=0.48\textwidth]{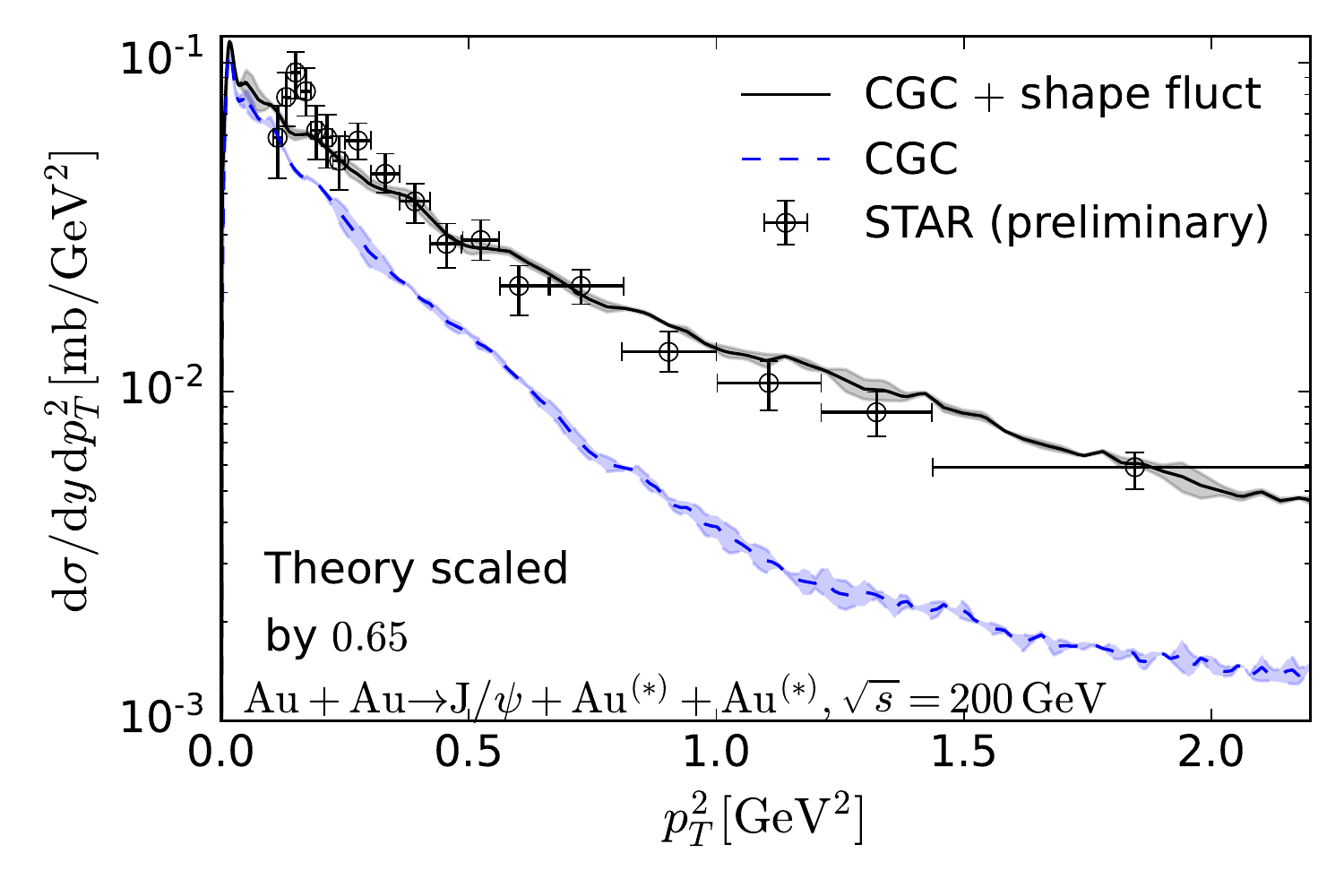}
          \caption{Incoherent \jpsi\ production cross sections in ultra peripheral Au$+$Au collisions at $\sqrt{s}=200\,\gev, y=0$. The preliminary STAR data is extracted from Ref.~\cite{star} for the total diffractive \jpsi production and is shown here  at $p_T^2>0.1\,\gev^2$ where the incoherent contribution dominates and the effect from the photon trasnverse momentum (neglected here) is very small.   }
         \label{fig:star_incoherent}
      \end{figure}

Next we study exclusive \jpsi production at larger meson $p_T$, where the incoherent contribution dominates. 
The preliminary STAR data, again extracted from Ref.~\cite{star}, and the incoherent cross section calculated from the CGC framework are shown in Fig.~\ref{fig:star_incoherent}.
The STAR data is only shown in the region where the incoherent channel dominates, and we again neglect the photon transverse momentum when calculating the incoherent cross section. 

In order to study the role of nucleon substructure fluctuations  in the $\xpom \sim 0.01$ region we calculate incoherent spectra with and without nucleon substructure fluctuations.
At $p_T^2 \gtrsim 0.3\,\gev^2$ the STAR data is found to clearly prefer the calculation with event-by-event fluctuating nucleon spatial structure, as without substructure the $p_T^2$ spectrum is too steep compared to the data. 
The enhancement at $p_T^2 \approx 0.2\,\gev^2$ can not be reproduced from our setup. We emphasize that the coherent cross section contributes only very little in this kinematics.

\section{Conclusions}
\label{sec:conclusions}

We have shown that the exclusive \jpsi production spectra at small squared meson transverse momentum as measured by the ALICE Collaboration suggest that non-linear saturation effects change the transverse density profile of the nucleus at small $x$. In particular, the ALICE \jpsi production data requires a steeper $p_T^2$ slope than what is suggested by the nuclear form factor.
When saturation effects are included in the calculation, an improved description of the experimentally measured $p_T^2$ spectra is obtained because non-linear dynamics changes the nuclear density profile. However, even with saturation effects we do not obtain a steep enough spectrum that would be compatible with the ALICE $\mathrm{Pb}+\mathrm{Pb} \to \jpsim + \mathrm{Pb}+\mathrm{Pb}$ data. Additionally, we find that the $p_T^2$ integrated \jpsi photoproduction data from both RHIC and LHC seems to indicate a stronger nuclear suppression than what is obtained from the applied CGC setup.

In addition to saturation effects, we quantify the role of two commonly neglected contributions: the interference effect (both nuclei can act as photon emitters) and the fact that the photons have non-zero transverse momenta in ultra peripheral collisions. At the LHC we predict a stronger suppression of the cross section at very small $p_T^2$, caused by interference effects, compared to what is visible in the ALICE data. 
The photon transverse momentum, on the other hand, is important around the diffractive minima. 
Only when both of these effects are included a good description of the $p_T^2$ differential cross section measured by the LHCb Collaboration is obtained. 
These contributions on the other hand have only a few percent effect on the $p_T^2$ integrated cross sections at LHC energies.

The fact that the measured \jpsi spectra in UPCs is more steeply falling than what is obtained from the applied CGC setup can  be explained 
if the gluonic size of the nucleus is larger than suggested by standard Woods-Saxon model parametrizations. The nuclear size parameter in the Woods-Saxon model is determined from low-energy electromagnetic interactions and can in principle differ from the gluonic size. Similar conclusions have been made based on exclusive $\pi^+ \pi^-$ production at RHIC~\cite{STAR:2022wfe}.
On the other hand, using the standard size of the Pb nucleus the calculated $p_T^2$ spectra in $\gamma + \mathrm{Pb} \to \jpsim + \mathrm{Pb}$ is compatible with the ALICE data except in the lowest $|t|$ bin. This shows that the method used by the ALICE Collaboration to remove the photon transverse momentum contribution and the interference effect from the measured \jpsi production spectra in UPCs does not exactly match with the approach taken in this work. In particular, with a non-zero photon transverse momentum in UPCs we obtain a less steeply falling spectrum in $\mathrm{Pb}+\mathrm{Pb}$ events compared to $\gamma+\mathrm{Pb}$, in contrast to ALICE data.

We also studied the role of nucleon substructure fluctuations in UPCs. Although the coherent cross section is only sensitive to the average interaction, there is larger nuclear suppression when substructure fluctuations are included due to larger local density variations in nuclei, and the non-linear nature of the effect. With substructure, one can generate very dense regions (along with more dilute spots in other regions), where many hot spots overlap, and saturation effects are greater there. Consequently the fluctuating nucleon substructure has a $\sim 7\%$ effect on the coherent cross section at midrapidity in the LHC kinematics. Compared to earlier implementations of the fluctuating nucleon geometry we here by construction obtain the same cross sections in $\gamma^*+p$ collisions in HERA kinematics with and without fluctuations, which allows for realistic predictions in LHC kinematics.  Even with substructure we do not find enough nuclear suppression in order to simultaneously describe overall normalization of the \jpsi photoproduction cross section in both $\gamma+p$ and $\gamma+$Pb collisions.

 If the normalization uncertainty is removed by studying the incoherent-to-coherent cross section ratio, the ALICE data clearly prefers nucleon substructure. The rapidity and center-of-mass-energy dependence of this ratio is sensitive to the geometry evolution. At large $p_T^2$ the STAR $p_T$ differential cross section measurement prefers the fluctuating nucleon substructure as well, and similarly large effects are predicted for LHC kinematics.

The future exclusive \jpsi spectra measurements from LHC experiments also in the (relatively) high-$p_T^2$ region as well as precise future EIC data, where the kinematics can be completely determined, will allow for a precise determination of the role of saturation effects on nuclear geometry, and on the event-by-event fluctuating nucleon structure in a nuclear environment.

\section*{Acknowledgments}
We thank J. Contreras, S. Klein, T. Lappi, L. Lavicka, J. Penttala, S. Räsänen and W. B. Schmidke for discussions. 
H.M. is supported by the Academy of Finland, the Centre of Excellence in Quark Matter, and projects  338263 and 346567, and under the European Union’s Horizon 2020 research and innovation programme by the European Research Council (ERC, grant agreement No. ERC-2018-ADG-835105 YoctoLHC) and by the STRONG-2020 project (grant agreement No. 824093). The content of this article does not reflect the official opinion of the European Union and responsibility for the information and views expressed therein lies entirely with the authors. F.S. is supported by the National Science Foundation under grant No. PHY-1945471, and partially supported by the UC Southern California Hub, with funding from the UC National Laboratories division of the University of California Office of the President. B.P.S. is supported by the U.S. Department of Energy, Office of Science, Office of Nuclear Physics under DOE Contract No.~DE-SC0012704.
Computing resources from CSC – IT Center for Science in Espoo, Finland and from the Finnish Grid and Cloud Infrastructure (persistent identifier \texttt{urn:nbn:fi:research-infras-2016072533}) were used in this work.

\appendix

\section{UPC scattering amplitude in coordinate space}
\label{appendix:amplitude}

We follow the formalism established in Ref.\, \cite{Xing:2020hwh} for the joint impact parameter and transverse momentum dependent cross sections. The differential cross section for exclusive vector meson production in ultra peripheral collisions is given by
\begin{align}
    \frac{\der \sigma^{A_1 + A_2 \to V_1 + A_1 + A_2}}{\der \pt^2 \der y }= \frac{1}{4\pi}  \int_{|\Bt|>B_{\mathrm{ min}}} \!\!\!\!\!\!\!\!\!\!\!\!\!\! \der^2 \Bt |\langle \Mcal^j(y,\pt,\Bt) \rangle_{\Omega}|^2 \,,
\end{align}
where the amplitude $\Mcal^j$ is given by the following convolution of the electric photon field $\Fcal^j$ and the vector meson production amplitude $\Acal$:
\begin{widetext}
\begin{align}
    \langle \Mcal^j(y,\pt,\Bt)  \rangle_{\Omega} = &\int \der^2 \Deltat \delta^{(2)}(\pt - \Deltat- \kt) \int \frac{\der^2 \kt}{(2\pi)^2}  \Big [  \langle \Acal(y,\Deltat) \rangle_{\Omega} \Fcal^j(y,\kt)  e^{-i\Bt \cdot \kt} +    \langle \Acal(-y,\Deltat) \rangle_{\Omega} \Fcal^j(-y,\kt) e^{-i\Bt \cdot \Deltat}  \Big ] \,. \label{eq:Mconv}
\end{align}
\end{widetext}
Here $\kt$ is the transverse momentum of the photon and $\Deltat$ is the momentum transfer from the gluon field. The first term in square brackets corresponds to the case in which the photon is emitted by nucleus 1 and scatters off nucleus 2, while the second term corresponds to the reverse case. The sub-amplitudes read
\begin{align}
    -i\Acal(y,\Deltat)  = &\int \der^2 \bt \ e^{-i \Deltat \cdot \bt} \int \der^2 \rt  \int_0^1 \frac{\der{z}}{4\pi} \nonumber \\
    & \times [\Psi_V^* \Psi_\gamma](Q^2,\rt,z) N(\rt,\bt,z,\xpom) \,,
\end{align}
where $\xpom = (M_V/\sqrt{s})e^{- y}$, and 
\begin{align}
    \Fcal^j(y,\kt) = 2Z {\aem}^{1/2} \kt^j \left[ \frac{F(\kt^2 + \xpom^2 M_p^2)}{\kt^2 + \xpom^2 M_p^2} \right] \label{eq:FYT} \,.
\end{align}
Here $F(\vec{k}^2)$ is the form factor of the nucleus, i.e. the Fourier transform of the thickness function, and $\xpom = \omega/(M_p \gamma) = (M_V/\sqrt{s})e^{y}$ where $M_p$ is the nucleon mass.

It is convenient to re-write the amplitude in Eq.\,\eqref{eq:Mconv} in terms of coordinate space functions:
\begin{multline}
    \Mcal^j(y,\pt,\Bt) = \int \der^2 \bt e^{-i\pt\cdot\bt} \left[ \widetilde{\Acal}(y,\bt) \widetilde{\Fcal}^j(y,\bt-\Bt)\right. \\ \left. +  \widetilde{\Acal}(-y,\bt) \widetilde{\Fcal}^j(-y,\bt+\Bt) e^{-i\pt\cdot\Bt} \right] \,,
    \label{eq:Amplitude_coordinate-space}
    \end{multline}
where the sub-amplitudes in coordinate space read
\begin{align}
    -i\widetilde{\Acal}(y,\bt) = \int \der^2 \rt  \int_0^1 \frac{\der{z}}{4\pi}  & [\Psi_V^* \Psi_\gamma](Q^2,\rt,z) \nonumber \\
    &\times N(\rt,\bt,z,\xpom) \,,
\end{align}
and
\begin{align}
    \widetilde{\Fcal}^j(y,\Bt)  = 2Z {\aem}^{1/2} \int  \frac{\der^2 \kt}{(2\pi)^2} 
    e^{i \kt \cdot \Bt } \kt^j \left[ \frac{F(\kt^2 + \xpom^2 M_p^2)}{\kt^2 + \xpom^2 M_p^2} \right] 
    \label{eq:IVector} \,.
\end{align}
We can employ the following approximation
\begin{align}
    \widetilde{\Fcal}^j(y,\Bt) 
    & = \frac{Z {\aem}^{1/2} \omega}{\pi \gamma} \frac{i\Bt^j}{|\Bt|} K_1\left( \frac{\omega |\Bt|}{\gamma} \right) \,,
\end{align}
which amounts to treating the source of photons as a point particle (Gauss' law), i.e., $F(\kt^2 + \xpom^2 M_p^2)=1$. Then one can write the amplitude as in Eq.\,\eqref{eq:Mi_simple}.

An equivalent expression (after a change of variables in Eq.\,\eqref{eq:Amplitude_coordinate-space}) for the amplitude $\Mcal^j(y,\pt,\Bt)$ is given by
\begin{multline}
    \Mcal^j(y,\pt,\Bt) = \int \der^2 \bt e^{-i\pt\cdot\bt} \left[ \widetilde{\Acal}(y,\bt) \widetilde{\Fcal}^j(y,\bt-\Bt)\right. \\ \left. +  \widetilde{\Acal}(-y,\bt-\Bt) \widetilde{\Fcal}^j(-y,\bt) \right] \,,
    \label{eq:Amplitude_coordinate-space2}
    \end{multline}
where the symmetry in the exchange between photon source and target is manifestly seen as $\Acal \leftrightarrow \Fcal$ with $y \to -y$. However, for numerical evaluation, it is more convenient to use Eq.\,\eqref{eq:Amplitude_coordinate-space} or Eq.\,\eqref{eq:Mi_simple}.

\section{Interference effect as a function of collision energy and momentum transfer}
\label{appendix:interference}

In order to illustrate the role of the interference effect on exclusive \jpsi production in ultra peripheral collisions at RHIC and at the LHC at midrapidity (where the interference is maximally large)  we define an effective photon flux $\tilde N$ which includes the interference effect:
\begin{equation}
    \tilde N(\omega, \pt^2) = \int \dd[2]{\Bt} n(\omega,\Bt) \left[ 1 - \cos (\pt \cdot \Bt) \right] \theta(|\Bt|-2R_{A}).
\end{equation}
Here the $\Bt$ dependent flux $n(\omega, \Bt)$ is given in Eq.~\eqref{eq:flux}. 
This is the effective flux factor that we encounter when calculating exclusive \jpsi production at midrapidity neglecting the photon transverse momentum, see  Eq.~\eqref{eq:interference_only_midrapidity}. In the following, we compare this flux to the vector meson momentum independent equivalent photon flux from a single ultrarelativistic nucleus $N(\omega)$ shown in Eq.~\eqref{eq:bint_flux}.

\begin{figure}
    \centering
         \includegraphics[width=0.5\textwidth]{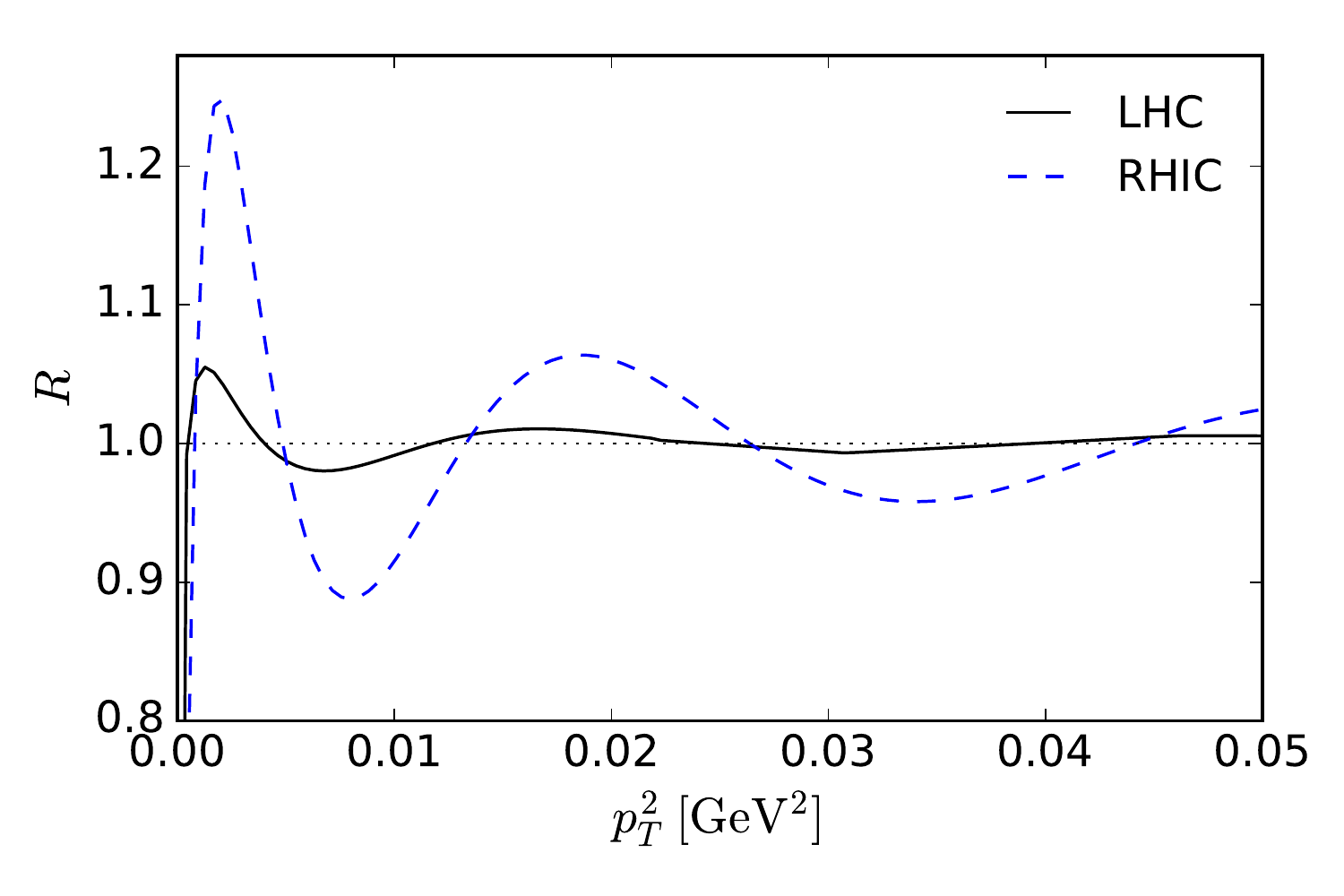}
          \caption{Interference effect as a function of \jpsi squared transverse momentum  $p_T^2$ at RHIC ($\sqrt{s}=200$ GeV) and at the LHC ($\sqrt{s}=5020$ GeV)}
         \label{fig:interference_effect}

    \end{figure}

Specifically, the role of the interference effect is demonstrated by showing in Fig.~\ref{fig:interference_effect} the ratio 
\begin{equation}
\label{eq:R}
    R = \frac{\tilde N(\omega,p_T^2)}{N(\omega)}.
\end{equation}
The ratio as a function of $p_T^2$ is shown separately for Pb$+$Pb collisions at $\sqrt{s}=5020\,\gev$ and for Au$+$Au collisions at $\sqrt{s}=200\,\gev$, corresponding to the LHC and RHIC kinematics, respectively. 

As  discussed e.g.~in Ref.~\cite{Bertulani:2005ru}, the interference effect is significant especially in the RHIC kinematics in the studied low momentum transfer region. At high $p_T^2$ the cosine term oscillates rapidly and the interference effect disappears in any realistic $p_T^2$ bin. At the LHC the interference effect has a numerically smaller contribution, but it may still change the cross section by $\sim 5\%$ at low but finite $\pt$, which is of the same order as the uncertainties in the LHC data, see discussion in Sec.~\ref{sec:lhc_spectra}. The interference effect also results in the cross section vanishing exactly at $p_T=0$ at midrapidity.

\bibliographystyle{JHEP-2modlong.bst}
\bibliography{refs}

\end{document}